\documentclass[12pt,preprint]{elsarticle/elsarticle}

\usepackage{amsmath}
\usepackage{amssymb}
\usepackage{multirow}
\usepackage[separate-uncertainty]{siunitx}
\usepackage{epsfig}
\usepackage{hyperref}
\usepackage{rotating}
\usepackage{color}
\usepackage{bm}
\usepackage[labelformat=simple]{subfig}
\usepackage[margin=1.1in]{geometry}

\usepackage{xspace}

\newcommand{\R} {{\sl R }}

\def\Iin     {\ifmmode I_{\mathrm{in}}                 \else $I_{\mathrm{in}}$                 \fi}
\def\Vin     {\ifmmode V_{\mathrm{in}}                 \else $V_{\mathrm{in}}$                 \fi}
\def\Vout    {\ifmmode V_{\mathrm{out}}                \else $V_{\mathrm{out}}$                \fi}
\def\VoutOne {\ifmmode V_{\mathrm{out}}^{\mathrm{(1)}} \else $V_{\mathrm{out}}^{\mathrm{(1)}}$ \fi}
\def\VoutTwo {\ifmmode V_{\mathrm{out}}^{\mathrm{(2)}} \else $V_{\mathrm{out}}^{\mathrm{(2)}}$ \fi}
\def\Iout    {\ifmmode I_{\mathrm{out}}                \else $I_{\mathrm{out}}$                \fi}

\def\amp {\ifmmode {\mathrm{A}} \else ${\mathrm{A}}$\fi}
\def\kG  {\ifmmode {\mathrm{kG}} \else ${\mathrm{kG}}$\fi}

\newcommand{\GeV} {\mathrm{GeV}}
\newcommand{\TeV} {\mathrm{TeV}}

\def\urltilde{\kern -.15em\lower .7ex\hbox{\~{}}\kern .04em}
\def\urldot{\kern -.10em.\kern -.10em}
\def\urlhttp{http\kern -.10em\lower -.1ex\hbox{:}\kern -.12em\lower 0ex\hbox{/}\kern -.18em\lower 0ex\hbox{/}}

\def\urltilde{\kern -.15em\lower .7ex\hbox{\~{}}\kern .04em}
\def\urldot{\kern -.10em.\kern -.10em}
\def\urlhttp{http\kern -.10em\lower -.1ex\hbox{:}\kern -.12em\lower 0ex\hbox{/}\kern -.18em\lower 0ex\hbox{/}}

\def\mev  {\ifmmode {\rm MeV} \else MeV\fi}
\def\mevc {\ifmmode {\rm MeV}/c \else MeV$/c$\fi}
\def\mevcc {\ifmmode {\rm MeV}/c^2 \else MeV$/c^2$\fi}
\def\gevc {\ifmmode {\rm GeV/c} \else GeV/c\fi}
\def\tevcc {\ifmmode {\rm TeV}/c^2 \else TeV$/c^2$\fi}
\def\tev   {\ifmmode {\rm TeV} \else TeV\fi}

\def\ol   {\overline}

\def\vtd  {\ifmmode |V_{td}| \else $|V_{td}|$\fi}
\def\vtb  {\ifmmode |V_{tb}| \else $|V_{tb}|$\fi}
\def\vts  {\ifmmode |V_{ts}| \else $|V_{ts}|$\fi}
\def\vcb  {\ifmmode |V_{cb}| \else $|V_{cb}|$\fi}

\newcommand{\Ds} {\ifmmode D_{\mbox{\sl s}}^{-}
                       \else $D_{\mbox{\sl s}}^{-}$\fi}
\newcommand{\Bs} {\ifmmode B_{\mbox{\sl s}}^{0}
                       \else $B_{\mbox{\sl s}}^{0}$\fi}
\newcommand{\Bsb} {\ifmmode \ol B_{\mbox{\sl s}}^{0}
                       \else $\ol B_{\mbox{\sl s}}^{0}$\fi}
\newcommand{\Bsh} {\ifmmode B_{\mbox{\sl s}}^H
                       \else $B_{\mbox{\sl s}}^H$\fi}
\newcommand{\Bsl} {\ifmmode B_{\mbox{\sl s}}^L
                       \else $B_{\mbox{\sl s}}^L$\fi}
\newcommand{\Dsl} {\ifmmode D_{\mbox{\sl s}}^{-} \ell^+
                       \else $D_{\mbox{\sl s}}^{-} \ell^+$\fi}
\newcommand{\xs} {\ifmmode x_{\mbox{\sl s}}
                       \else $x_{\mbox{\sl s}}$\fi}
\newcommand{\xd} {\ifmmode x_d \else $x_d$\fi}
\newcommand{\lxy} {\ifmmode L_{\rm xy} \else $L_{\rm xy}$\fi}
\newcommand{\dgam} {\ifmmode \Delta\Gamma \else $\Delta\Gamma$\fi}
\newcommand{\dm} {\ifmmode \Delta m \else $\Delta m$\fi}
\newcommand{\ctau} {\ifmmode c\tau \else $c\tau$\fi}

\newcommand{\enmath}[1]{\ensuremath{#1}\xspace}

\renewcommand{\R}          {\enmath{ \mathbb{R}  }}
\newcommand{\Nmom}         {\enmath{ \mathcal{N} }}
\newcommand{\intdz}        { \int\limits_0^\infty dz }

\renewcommand{\vec}[1]     {\enmath{ \tilde {\bm {#1}} }}
\newcommand{\vecl}[2]      {\enmath{ \tilde {\bm {#1}}_{#2} }}
\newcommand{\vecu}[2]      {\enmath{ \tilde {\bm {#1}}^{#2} }}
\newcommand{\veclu}[3]     {\enmath{ \tilde {\bm {#1}}_{#2}^{#3} }}

\newcommand{\op}[1]        {\enmath{ \hat   {\bm {#1}} }}
\newcommand{\opl}[2]       {\enmath{ \hat   {\bm {#1}}_{#2} }}
\newcommand{\opu}[2]       {\enmath{ \hat   {\bm {#1}}^{#2} }}
\newcommand{\oplu}[3]      {\enmath{ \hat   {\bm {#1}}_{#2}^{#3} }}

\newcommand{\I}[1]         { {\left(#1\right)} }

\newcommand{\Dkl}[2]       {\enmath{ D_\mathrm{KL}\left( #1\middle\|#2 \vphantom{\vec{f}}\right) }}
\newcommand{\tr}[1]        {\ensuremath{\mathrm{tr}\left(#1\right)}}
\newcommand{\dt}[2]        { \left(\vec{#1} - \vec{#2}\right) }
\newcommand{\repart}[1]    { \operatorname{Re}\left(#1\right) }

\newcommand{\ofz}          {\enmath{\left(z\right)}}
\newcommand{\ofzs}         {\enmath{\left(z^\star\right)}}
\newcommand{\ofzi}         {\enmath{\left(z_i\right)}}

\newcommand{\ofx}          {\enmath{\left(x\right)}}
\newcommand{\ofxt}         {\enmath{\left(x;\overline\theta\right)}}

\newcommand{\ofy}          {\enmath{\left(y\right)}}

\newcommand{\Mf}           {\enmath{  \opl{\mathcal M}{nm} }}
\newcommand{\Mfi}          {\enmath{  \oplu{\mathcal M}{~m}{i} }}

\newcommand{\Eb}[1]        {\enmath{ E_{#1} }}
\newcommand{\Fb}[1]        {\enmath{ F_{#1} }}

\newcommand{\inner}[2]     {\enmath{ \left\langle #1, #2 \right\rangle }}

\newcommand{\innerEE}[2]   {\enmath{ \inner{ E_{#1} }{ E_{#2} }  }}

\newcommand{\innerFF}[2]   {\enmath{ \inner{ F_{#1} }{ F_{#2} }  }}

\journal{Nuclear Instruments and Methods A}

\begin{document}

\title{\textbf{Functional Decomposition}\\ A new method for search and limit setting.}

\author[um]{R.~Edgar\corref{cor1}}
\ead{edgarr@umich.edu}

\author[um]{D.~Amidei}
\ead{amidei@umich.edu}

\author[um]{C.~Grud}
\ead{cgrud@umich.edu}

\author[um]{K.~Sekhon}
\ead{ksekhon@umich.edu}

\cortext[cor1]{Corresponding author}
\address[um]{University of Michigan, Ann Arbor, \\
450 Church St. Ann Arbor, MI 48109, United States}


\begin{abstract}
In the analysis of High-Energy Physics data, it is frequently desired to separate resonant signals from a smooth, non-resonant background.
This paper introduces a new technique - functional decomposition (FD) - to accomplish this task.
It is universal and readily able to describe often-problematic effects such as sculpting and trigger turn-ons.

 Functional decomposition models a dataset as a truncated series expansion in a complete set of orthonormal basis functions, using a process analogous to Fourier analysis.  A new family of orthonormal functions is presented, which has been expressly designed to accomplish this in a succinct way.  A consistent signal extraction methodology based on linear signal estimators is also detailed, as is an automated method for selecting the method's (few) hyperparameters and preventing over-fitting.

The full collection of algorithms described in this paper have been implemented in an easy-to-use software package, which will also be briefly described.

\end{abstract}

\maketitle

\section{Motivation and Overview}
A substantial body of searches for new physical phenomena are conducted under the resonance {\it ansatz}: new physics is presumed to present as a localized deviation (a resonance) from an otherwise-smooth background.  The smooth background is most commonly modeled using Monte Carlo simulation, data sidebands, parametric fits to functional forms, or some combination of these.  While all of these approaches are applied frequently and with success, each also has disadvantages: Monte Carlo requires careful testing and sophisticated control of systematic uncertainties, and can be computationally infeasible for very large datasets;  data sidebands are usually limited in statistical precision and often capture only some of the physical phenomena contributing to the region of interest; parametric fits rely on functions that are approximate or ad-hoc, and therefore may not adequately capture all features of the smooth background.

\begin{figure*}[!ptb]
\begin{center}
  \subfloat[]{
    \includegraphics[width=0.75\textwidth, clip]{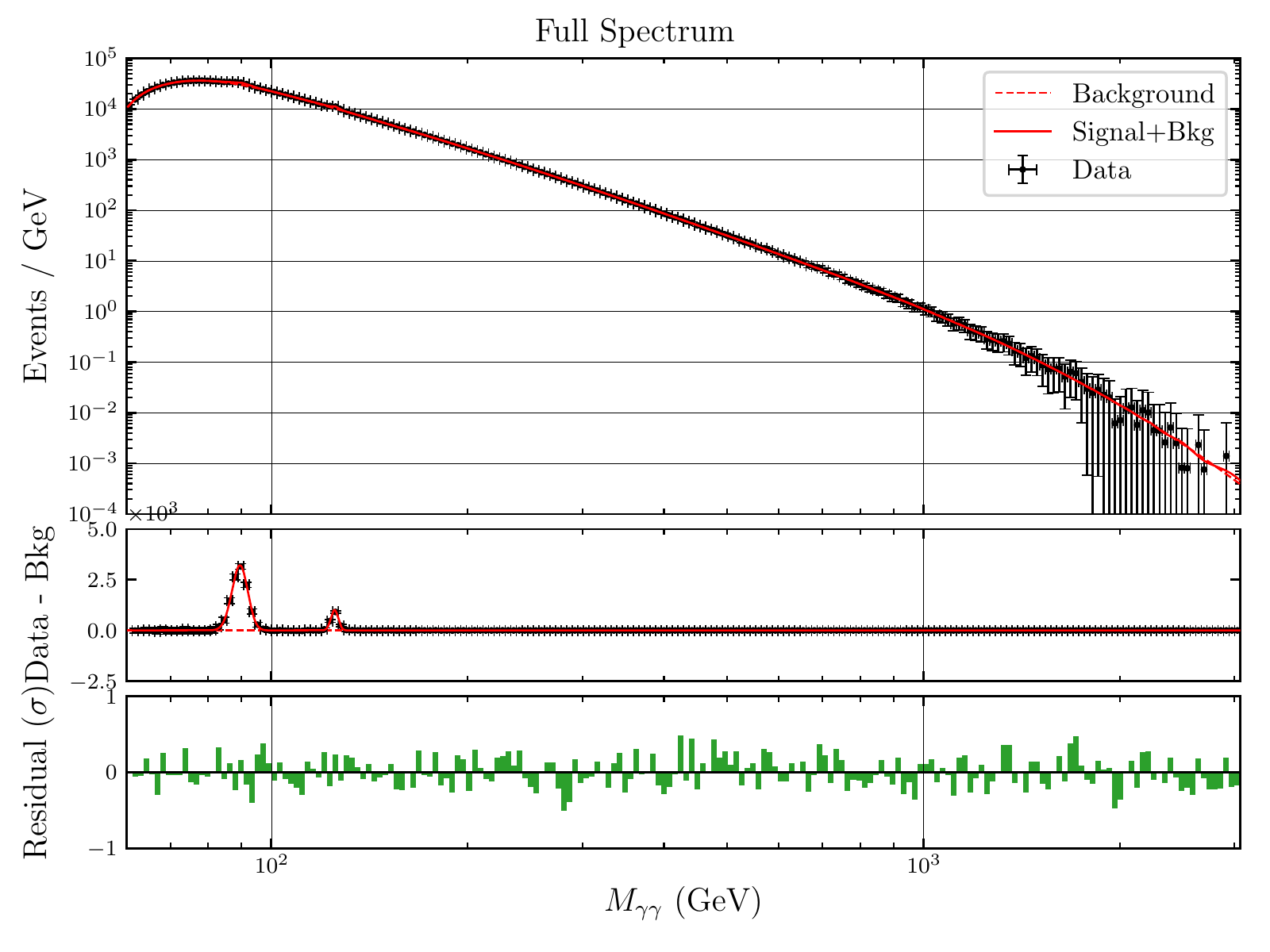}
    \label{fig:hook_full}
  }

  \subfloat[]{
    \includegraphics[width=0.75\textwidth, clip]{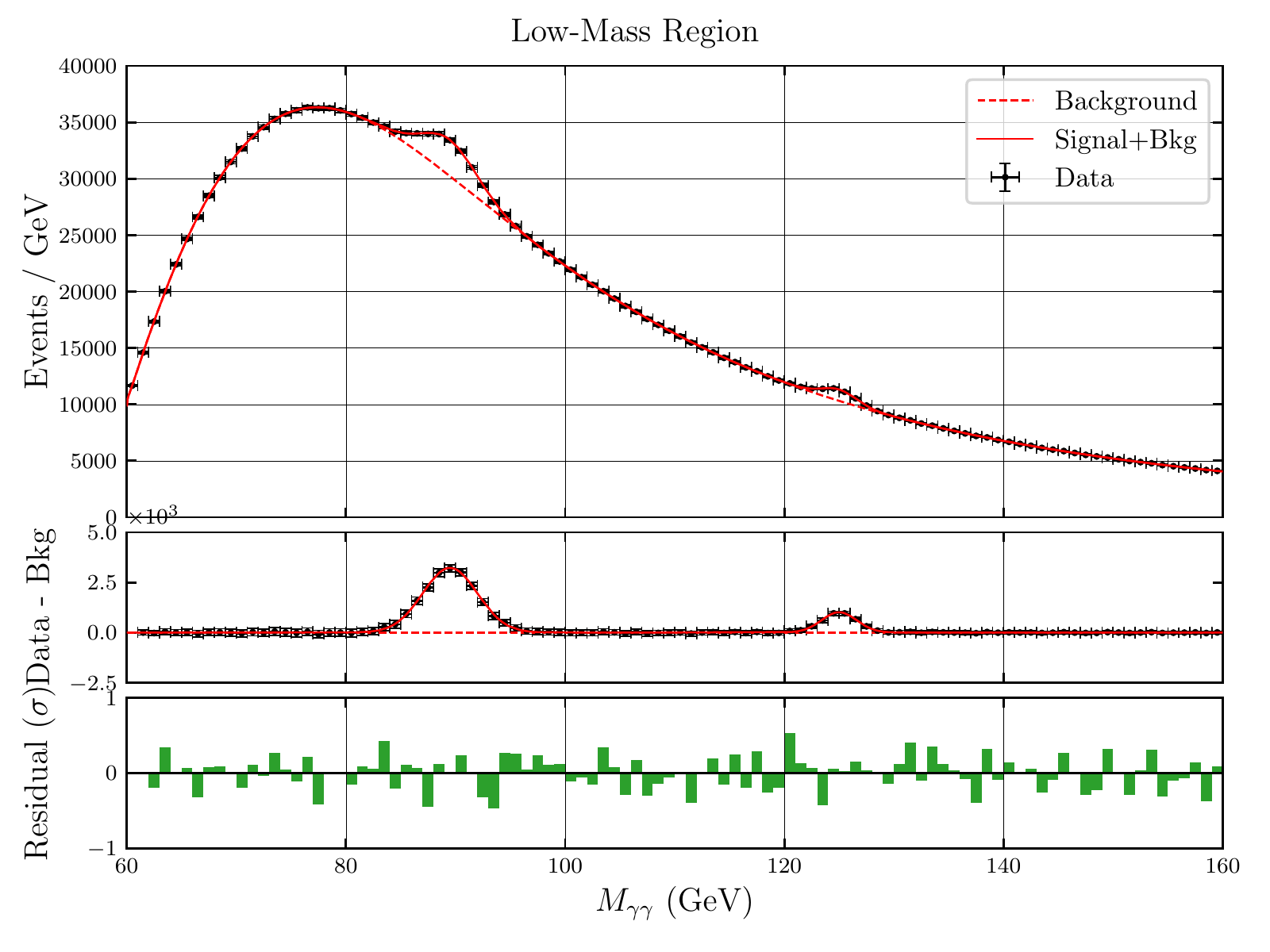}
    \label{fig:hook_lowmass}

  }
  \caption{Functional decomposition applied to a simulated data spectrum, designed to approximate the invariant mass $M_{\gamma\gamma}$ of the two-photon final state at $13~\TeV$ at CERN's LHC.  The simulated spectrum has roughly $5\times10^7$ events scaled down to $2\times10^6$ in order to approximate the statistically-asymptotic behavior. The figures show the decomposition plotted over the full range (Fig.~\protect\subref{fig:hook_full}) and only in the low-mass region (Fig.~\protect\subref{fig:hook_lowmass}).  Both figures show the same data and decomposition; only the $x$ axis ranges and the binning differ. \label{fig:hook}}
\end{center}
\end{figure*}

A resonance search technique that addresses these disadvantages is thus a tool with potentially widespread application.  Functional decomposition is such a tool.  Its power can be seen in Figure~\ref{fig:hook}, which illustrates the application of FD to a particularly troublesome case: the production of a purely data-driven model for a spectrum that includes both a turn-on and two known resonant peaks.  The performance on this particular example will be explored in more detail during the course of this paper, but the excellent modeling of the data over its full range is evident, including the turn-on, peak, resonances and tail.

Functional Decomposition is a form of Orthogonal Density Estimation (ODE;~\cite{Efromovich2010}).  It operates by applying a transformation to the variable of interest and modeling the resulting dimensionless variable using a complete set of orthonormal functions.  Judicious choices of the basis and transformation ensure that the smooth background has a succinct representation, using only the first few terms in the series.  The remaining (i.e. higher-order) terms are then available to construct estimators for the resonant contributions.

This model can be written as:
\begin{align}
  z          &= T\ofxt                                                                        \label{eq:Xfrm} \\
  \Omega\ofz &= \sum\limits_{n=0}^{\Nmom-1} c^n E_n\ofz + \sum\limits_{m=0}^{N_s} s^\I{m}S_\I{m}\ofz \label{eq:Xmdl} \\
  \Omega\ofx &= \sum\limits_{n=0}^{\Nmom-1} c^n \left(\frac{dz}{dx}\right)E_n\ofz + \sum\limits_{m=0}^{N_s} s^\I{m}S_\I{m}\ofx \label{eq:Xmdlx}\, .
\end{align}
Here, $x$ is the initial variable of interest and $z=T\ofxt$ is the corresponding dimensionless variable. The transformation $T$ is parameterized by the vector $\overline\theta$ (the transformation hyperparameters).  The functions $\left\{E_n\right\}$ are a complete set of orthonormal functions, from which the first \Nmom are retained for the background model.  The parameters $c_n$ are the coefficients of the background distribution, presumed zero if $n\geq\Nmom$. Finally, $S_m\ofz$ are some number $N_s$ of resonant contributions, each of which has a corresponding normalization $s_m$, and each of whose shapes are presumed to be known.

The use of a complete, orthonormal set of basis functions guarantees that any smooth function can be described by such a series expansion.  But the performance is entirely determined by the particular choice of basis and transformation; succinct expansions (that is, \Nmom is small) retain maximum information for estimating the resonant contributions, while more verbose choices reduce the resonant sensitivity and often produce approximations that are not positive-definite (and therefore are not valid probability distributions).

The remainder of this paper is devoted to establishing particulars for this technique that are effective for resonance searches and measurements in high-energy physics.  It is organized as follows:
\begin{itemize}
\item Section~\ref{sec:test_spectrum}: The test spectrum;
\item Section~\ref{sec:orthexp}: The orthonormal exponentials: a basis for falling spectra;
\item Section~\ref{sec:xfrm}: The power-law transformation;
\item Section~\ref{sec:decomp}: Decomposing the dataset;
\item Section~\ref{sec:sb_est}: Estimating signal and background parameters;
\item Section~\ref{sec:hyper_opt}: Optimization of hyperparameters;
\item Section~\ref{sec:statistics}: Statistical interpretation;
\item Section~\ref{sec:summary}: Summary and discussion; the FD package.
\end{itemize}

Several mathematical results will be required along the way.  The proofs and derivations are relegated to appendices to avoid some otherwise-lengthy digressions.

\subsection{Notation}
For convenience and consistency, the following notational conventions will be used throughout:
\begin{itemize}
  \item A function is written $f\ofz$;
  \item its vector representation in a Hilbert space is denoted \vec{f};
  \item and the individual components of that vector are \vecl{f}{n}.
  \item Similarly, operators are written \op{O} with components \opl{O}{nm}.  The operator form of $f\left(z\right)$ is written as \op{f}.
  \item Families of functions have parenthesized indices (i.e.  $f_\I{n}\ofz$; \vecl{f}{\I n}; \vecl{f}{\I ni}).  This distinguishes them from the Hilbert-space indices.
  \item Einstein summation convention is used whenever possible: repeated indices with one index lowered and the other raised imply summation (i.e. $c_id^i =\sum\limits_i c_i d_i$).  Explicit sums are used when required for clarity or to indicate summation limits.
  \item Raised and lowered indices denote the same numerical values; the position is used only to indicate implied summation.
  \item Multiplications written $\op{O}\vec{f}$ are to be read as $\oplu{O}{n}{~m}\vecl{f}{m}$ (matrix multiplication).
  \item Angle brackets are sometimes used for inner products: $\inner{u}{v} = \vecl{u}{n}\vecu{v}{n}$.
\end{itemize}

\section{The test spectrum}\label{sec:test_spectrum}
FD will be illustrated with the aid of an example spectrum, already shown in Figure~\ref{fig:hook}.  This spectrum was constructed to exhibit several features that are usually the source of some difficulty: a large mass range and an event rate that spans some six orders of magnitude, a turn-on in the low-mass region, and two known resonant peaks which, for the purpose of conducting a search, must be included as part of the background.

The test spectrum was designed with particular reference to the two-photon final state in $pp$ collisions at $13~\TeV$ at CERN's LHC.  Forming the invariant mass $M_{\gamma\gamma}$, one expects a smooth, high-statistics background with one resonant peak from the Higgs boson and a second resonant peak from the $Z$.  Though the $Z$ does not decay to two photons, $Z\rightarrow ee$ decays are occasionally misidentified as two-photon events.  This can lead to a substantial `fake' $Z$ contribution when the production rate is sufficiently high, as is the case at the LHC.  At low mass, the spectrum is modified by the trigger and selection thresholds.

The model for the continuum $M_{\gamma\gamma}$ background consists of $5\times10^7$ events generated according to the probability distribution
\begin{equation}
  P\ofx = p_5G\ofx + p_0\left(1-y\right)^{p_1} y^{ p_2 + p_3\log y + p_4\log^2 y }
\end{equation}
where
\begin{align}
  G\ofx &= \frac{1}{\sqrt{2\pi}p_7}\exp\left[ -\frac{1}{2} \left(\frac{x-p_6}{p_7}\right)^2 \right] \\
  y     &= \frac{x}{13~\TeV} \, ,
\end{align}
and the constant parameters are given by
\begin{align*}
  p_0 &=  \hphantom{-} \num{9.50700e4} \\
  p_1 &=  \hphantom{-} \num{4.58242e1} \\
  p_2 &=              -\num{1.21268e1} \\
  p_3 &=              -\num{1.51309e0} \\
  p_5 &=  \hphantom{-} \num{2.38849e-1} \\
  p_6 &=              -\num{8.38068e6} \\
  p_7 &=  \hphantom{-} \num{3.34980e1} \\
  p_8 &=  \hphantom{-} \num{1.83062e1} \, .
\end{align*}
The probability distribution $P\ofx$ is one of the well-known `dijet functions', typically used in resonance searches involving the strong interaction (e.g.~\cite{PhysRevD.55.R5263},~\cite{CDF:2009DijetSearch},~\cite{CMS-PAS-EXO-16-056},~\cite{dijet2017}), with the addition of a Gaussian term to produce the turn-on.
The $Z$ contribution is modeled using $5\times10^5$ normally-distributed events with a mass of $89.5~\GeV$ and a width of $2.5~\GeV$.  The Higgs is modeled using $1\times10^5$ normally-distributed events with a mass of $125.0~\GeV$ and a width of $1.6~\GeV$.  The widths correspond to the natural widths of the particles modified by typical experimental resolutions.  In the case of the $Z$, the resolution is broadened and the mass is shifted downward by misidentification of electrons as photons.

The test sample is scaled down by a factor of $25$, for a total of $2\times10^6$ background events, $2\times10^4$ $Z$ events, and $4\times10^3$ Higgs events.  This scaling is applied to ensure that any biases or spurious signals introduced by the method are clearly visible and not obscured by statistical fluctuations.

\section{The orthonormal exponentials: a basis for falling spectra}\label{sec:orthexp}
\begin{figure*}[!tb]
\begin{center}
  \includegraphics[width=\textwidth, clip]{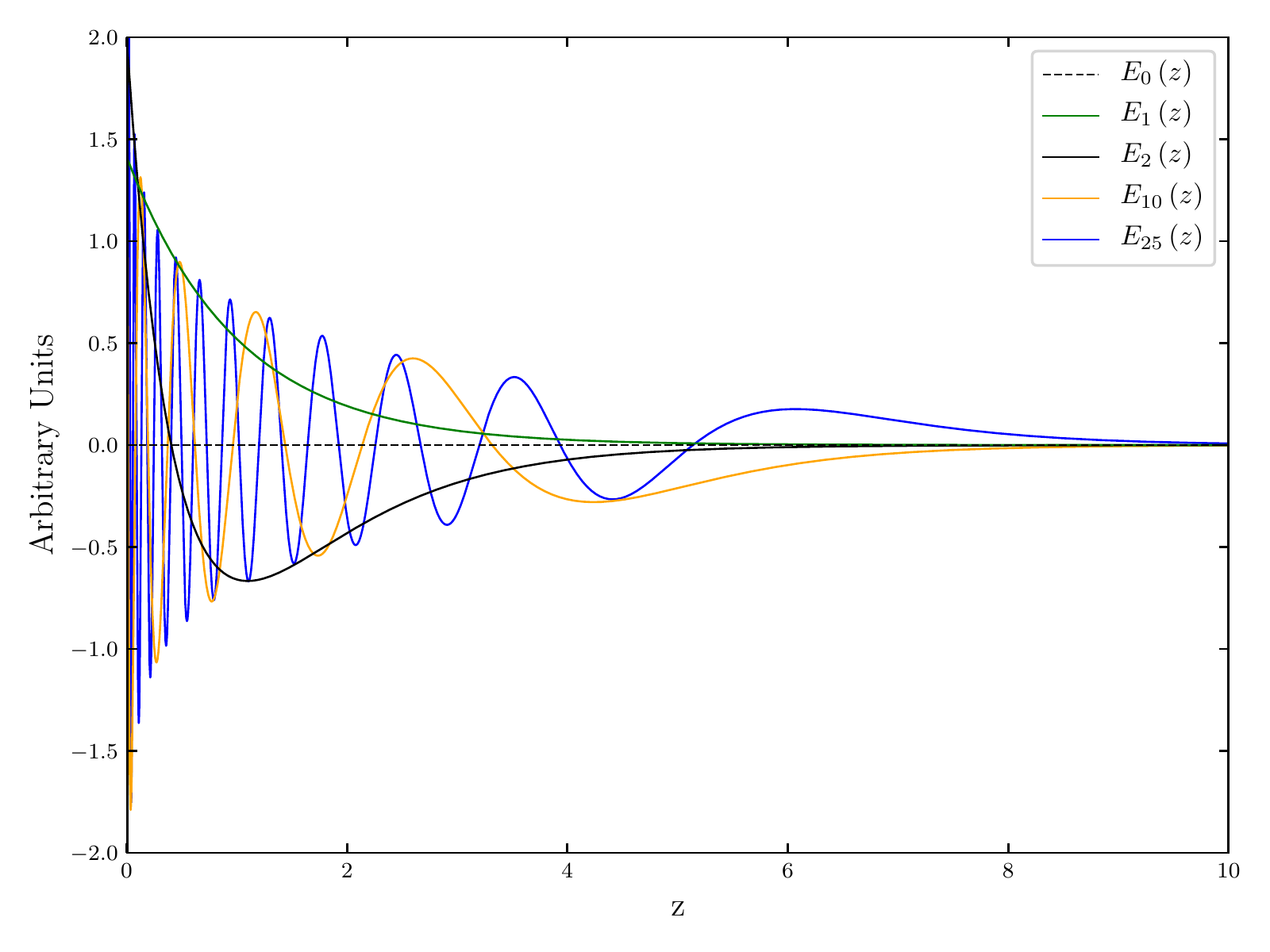}
  \caption{A selected few of the orthonormal exponentials. \label{fig:Efuncs}}
\end{center}
\end{figure*}

Existing sets of orthonormal functions (orthogonal polynomials, Bessel functions, trigonometric functions, etc) do not generally produce succinct representations of exponentially-falling spectra.  To address this, a new set of orthonormal functions has been constructed from the exponential function.
Some previous efforts to construct orthonormal bases from finite sets of exponential functions have been made in the field of signal processing (\cite{Kautz1954},~\cite{Ross1964},~\cite{Lai1985}).  None, to the authors' knowledge, describe the infinite family of functions detailed here.

The choice of the exponential function is motivated by several considerations.  Qualitatively, the tails of many spectra can often be approximated using members of the exponential family (simple exponentials or exponentials of a polynomial, for example).  Furthermore, the simple exponential is an {\it entropy-maximizing distribution}.  That is, in an information-theoretic sense, the exponential is the least-informative (most delocalized) distribution of all distributions sharing the same mean.  This intuitively fits the resonance {\it ansatz} of smoothly-falling delocalized backgrounds and localized signals.

We begin by defining a non-orthogonal family of functions (the \textit{exponential basis}) coupled with the $L^2$ inner product:
\begin{align}
  \Fb{n}\ofz    &= \sqrt{2}e^{-nz} \label{eq:nonorthodef}\\
  \inner{f}{g}  &= \intdz f\ofz g\ofz \, .
\end{align}
Here $z$ is the transformed variable as above (Eq.~\ref{eq:Xfrm}).  This set of functions is complete with respect to continuous probability distributions defined on  $\left[0,\infty\right)$ (see \ref{sec:completeness} for proof).  The \textit{orthonormal exponentials} are then defined in terms of this inner product and the exponential basis functions:
\begin{align}
  \Eb{n}\ofz     &= \sum\limits_{m=1}^n\opl{d}{nm} \Fb{m}\ofz \label{eq:ortho_def} \\
  \innerEE{n}{m} &=\delta_{nm}\, .
\end{align}
A selected few of the orthonormal exponentials can be seen in Figure~\ref{fig:Efuncs}.

The coefficients \opl{d}{nm} can be derived numerically using any number of well-known methods (e.g. Gram-Schmidt).  However, the inner product matrix $\innerFF{n}{m}$ is ill-conditioned; this constrains the usefulness of numerical solutions to just the first few orthonormal exponentials.  The authors have therefore derived an exact solution for the coefficients \opl{d}{nm}, as well as recurrence relations for the functions (see \ref{sec:orthcoeffs}).  These recurrence relations take the form
\begin{align}
  \Eb{1}\ofz   &= \sqrt{2} e^{-z} \\
  \Eb{n+1}\ofz &= \frac{1}{\phi_{2n+1}}\left(
                               4e^{-z}\Eb{n}\ofz 
                             - \frac{2}{\phi^2_{2n}}\Eb{n}\ofz
                             - \phi_{2n-1} \Eb{n-1}\ofz
                  \right) \\
  \phi_n       &= \sqrt{1-\frac{1}{n^2}} \label{eq:recurrence_first} \, .
\end{align}
This provides a fast and numerically stable method for evaluating the orthonormal exponentials.  Moreover, evaluating $\Eb{N}\ofz$ for some $z$ with the recurrence relations naturally produces $\Eb{n}\ofz$ for all $n<N$, which is quite advantageous for the present application.

\section{The power-law transformation}\label{sec:xfrm}
We next specify the transform $z=T\left(x;\overline\theta\right)$.  There are few constraints on the choice of this transformation.  It must (invertibly) map the range $\left[x_0,\infty\right)$ of the variable of interest to $\left[0,\infty\right)$ while rendering it dimensionless.  It must be continuous.  It is desirable that it have some flexibility that can be applied to ensuring that the resulting decompositions are succinct; on the other hand, too many free parameters become difficult to handle.

We find that the power-law transformation
\begin{equation}
  z = \left(\frac{x-x_0}{\lambda}\right)^\alpha \label{eq:Xpowerlaw}
\end{equation}
meets these requirements well.  There are three hyperparameters: $x_0$ specifies the start of the distribution, $\lambda$ is a positive scale parameter, and $\alpha$ is a positive, dimensionless exponent.  Intuitively, the hyperparameters adjust the shape of the tail (all orthonormal exponentials approach $e^{-z}$ as $z\rightarrow\infty$) as well as the spacing of the different degrees of freedom across the spectrum.

Because every choice of $\lambda$ and $\alpha$ produces a distinct (but still complete) orthonormal basis, they are, in a certain sense, arbitrary.  But as stated above, careful selection of their values can greatly affect the number of terms required to model the background.  Optimal selection of the hyperparameters is thus crucial to ensuring FD's efficacy.  But the optimization of the hyperparameters must be performed numerically, and it is necessary to recompute the series coefficients $c_n$ at each iteration of the optimization.

For large datasets, a from-scratch re-computation of the series coefficients can be expensive.  Luckily, there is another way.  Each choice of hyperparameters by design produces a distinct orthonormal basis on the same underlying Hilbert space.  Thus the decomposition \veclu{f}{n}{\star} of some function $f\ofx$ with hyperparameters $\theta^\star$ is connected by a linear transformation to the decomposition \vecl{f}{n} with hyperparameters $\theta$.  A general treatment of these transformation matrices is found in \ref{sec:xfrm_general}.

The power-law transformation has the nice property that the transformation matrices between different choices of hyperparameters are calculable:
\begin{equation}
  \vecl{f}{m}  = \exp\left[ c\op C + s\op S  \right] \veclu{f}{n}{\star}\, ,
\end{equation}
where $\exp$ is the matrix exponential and the matrices \op{C} and \op{S} are the mathematical constants
\begin{align}
  \opl{C}{nm} &= -\sum\limits_{i=1}^n\sum\limits_{j=1}^m \opl{d}{ni}\opl{d}{mj}\frac {i}{\left(i+j\right)^2} \left[ 1-\gamma-\ln\left(i+j\right) \right] \\
  \opl{S}{nm} &= -\sum\limits_{i=1}^n\sum\limits_{j=1}^m \opl{d}{ni}\opl{d}{mj}\frac {i}{\left(i+j\right)^2}
\end{align}
($\gamma=0.5772\dots$ is the Euler-Mascheroni constant).  The transformation parameters are
\begin{align}
  c      &= \ln\frac{\alpha}{\alpha^\star}                              \\
  s      &= -\frac{\alpha c}{e^c - 1} \ln\frac{\lambda}{\lambda^\star} \, .
\end{align}
See \ref{sec:xfrm_powerlaw} for a derivation of this result.  Being mathematical constants, the matrices \op{C} and \op{S} need be computed only once.  To apply a transformation in practice, then, requires only the computation of the action of a matrix exponential on the original decomposition \veclu{f}{n}{\star}.  This is a common enough operation that fast and efficient implementations exist in most widely-available numerical linear algebra libraries.

\section{Decomposing the dataset}\label{sec:decomp}
Given some dataset $\left\{x_i\right\}$ with $M$ measurements, the customary approach to obtaining the parameters \vecl{f}{n} is to choose them to maximize the log-likelihood of the data.  This approach suffers in performance when the number of parameters is large and, if unbinned, for large datasets.  Moreover, the estimate for the parameters will change depending on the model - that is, \vecl{f}{1} with $\Nmom=2$ will generally be different from \vecl{f}{1} with $\Nmom>2$.

The use of orthogonal series opens an additional avenue to estimate the parameters.  Supposing that $f\ofx$ is the underlying probability distribution, the parameters can be extracted with the inner product:
\begin{equation}
  \veclu{f}{n}{\mathrm{true}} = \inner{E_{n}}{f} = \intdz f\ofz E_n\ofz = \lim\limits_{M\rightarrow\infty} \frac{1}{M} \sum\limits_{i=1}^{M} E_n\ofzi \, ,
\end{equation}
where $z_i = z\left(x_i\right)$ and the last equality follows from the strong law of large numbers.  It then suffices to define the empirical moments as
\begin{equation}
  \vecl{f}{n} = \frac{1}{M} \sum\limits_{i=1}^{M} E_n\ofzi\, .
\end{equation}
This same result can also be obtained by direct computation, treating the data as a normalized ``comb'' of Dirac $\delta$-functions.   Furthermore, it can be shown to agree with the parameters estimated by maximizing the unbinned log-likelihood assuming infinitely-many free parameters \vecl{f}{n}.  The first few parameters for the test data spectrum are shown in Table~\ref{tab:testdata_decomp}.

\begin{table}[tb]
\begin{center}
\scriptsize
\def\arraystretch{1.35}
\begin{tabular}{rS[table-format=2.2e1]rS[table-format=2.2e1]rS[table-format=2.2e1]}
\hline\hline
  Moment       & \multicolumn{1}{r}{Value} & Moment & \multicolumn{1}{r}{Value} & Moment & \multicolumn{1}{r}{Value}  \\
  \hline
               &          & \vecl{f}{11} &  8.25e-3 & \vecl{f}{22} &  2.27e-3 \\
  \vecl{f}{ 1} &  5.34e-1 & \vecl{f}{12} &  3.09e-4 & \vecl{f}{23} & -3.96e-3 \\
  \vecl{f}{ 2} & -4.49e-1 & \vecl{f}{13} & -7.40e-3 & \vecl{f}{24} &  5.87e-4 \\
  \vecl{f}{ 3} & -9.84e-2 & \vecl{f}{14} &  3.34e-3 & \vecl{f}{25} &  3.30e-3 \\
  \vecl{f}{ 4} &  2.01e-1 & \vecl{f}{15} &  5.18e-3 & \vecl{f}{26} & -1.06e-4 \\
  \vecl{f}{ 5} & -2.03e-2 & \vecl{f}{16} & -6.16e-3 & \vecl{f}{27} & -3.18e-3 \\
  \vecl{f}{ 6} & -5.58e-2 & \vecl{f}{17} & -1.79e-3 & \vecl{f}{28} &  8.87e-4 \\
  \vecl{f}{ 7} &  1.47e-2 & \vecl{f}{18} &  6.70e-3 & \vecl{f}{29} &  3.17e-3 \\
  \vecl{f}{ 8} &  1.64e-3 & \vecl{f}{19} & -1.92e-3 & \vecl{f}{30} & -2.58e-3 \\
  \vecl{f}{ 9} & -1.07e-2 & \vecl{f}{20} & -4.88e-3 & \vecl{f}{31} & -2.12e-3 \\
  \vecl{f}{10} & -5.27e-3 & \vecl{f}{21} &  3.89e-3 & \vecl{f}{32} &  4.06e-3 \\
  \hline
  \hline
\end{tabular}
  \caption{The first $32$ moments of the test data spectrum.  These are computed with $\lambda=32.90~\GeV$ and $\alpha=0.60$.\label{tab:testdata_decomp}}
\end{center}
\end{table}

Asymptotically, the parameters \vecl{f}{n} are normally distributed regardless of the underlying function $f\ofz$ (see \ref{sec:meanvar}).
The uncertainties on the moments can therefore be represented by $\opl{\Sigma}{\vec{f}nm}/M$, where the covariance is given by
\begin{align}
  \opl{\Sigma}{\vec{f}nm} &= \frac{1}{M} \sum\limits_{i=1}^{M} E_n\ofzi E_m\ofzi - \vecl{f}{n}\vecl{f}{m} \label{eq:empcov} \\
                          &= \vecu{f}{i}\opl{I}{inm} - \vecl{f}{n}\vecl{f}{m}
\end{align}
and \opl{I}{inm} is the triple-product tensor $\opl{I}{inm}=\intdz E_i\ofz E_n\ofz E_m\ofz$.
This covariance matrix is exactly calculable from the decomposition \vecl{f}{n} (see \ref{sec:funcovfact}).  This allows fast computation of the covariance matrix in $\mathcal O\left(N^2\right)$ time, instead of the $\mathcal O\left(N^2M\right)$ required for a direct computation.  

Remarkably, the $N\times N$ covariance matrix of the first $N$ moments is a function of \textit{only} the first $2N$ moments.   This makes the dissemination of the covariance matrix at best a convenience - one need only report the first $2N$ moments to exactly capture the $N$ desired moments along with their covariance.  Even reporting the first $N$ moments alone is often enough - they capture their own covariance to a very good approximation!

\section{Estimating signal and background parameters}\label{sec:sb_est}
The infinite set of parameters \vec{f} from the previous section is exactly equivalent to the original dataset.  It can be regarded as \textit{being} the data, transformed to a different representation.  The next task is to extract the parameters $c_n$ and $s_m$ as per Eq.~\ref{eq:Xmdl}.

\begin{figure*}[!ptb]
\begin{center}
  \subfloat[A comparison between the signals and the corresponding minimum-variance estimators for the $h$ and $Z$ (assuming the covariance of the test sample).]{
    \includegraphics[width=0.75\textwidth, clip]{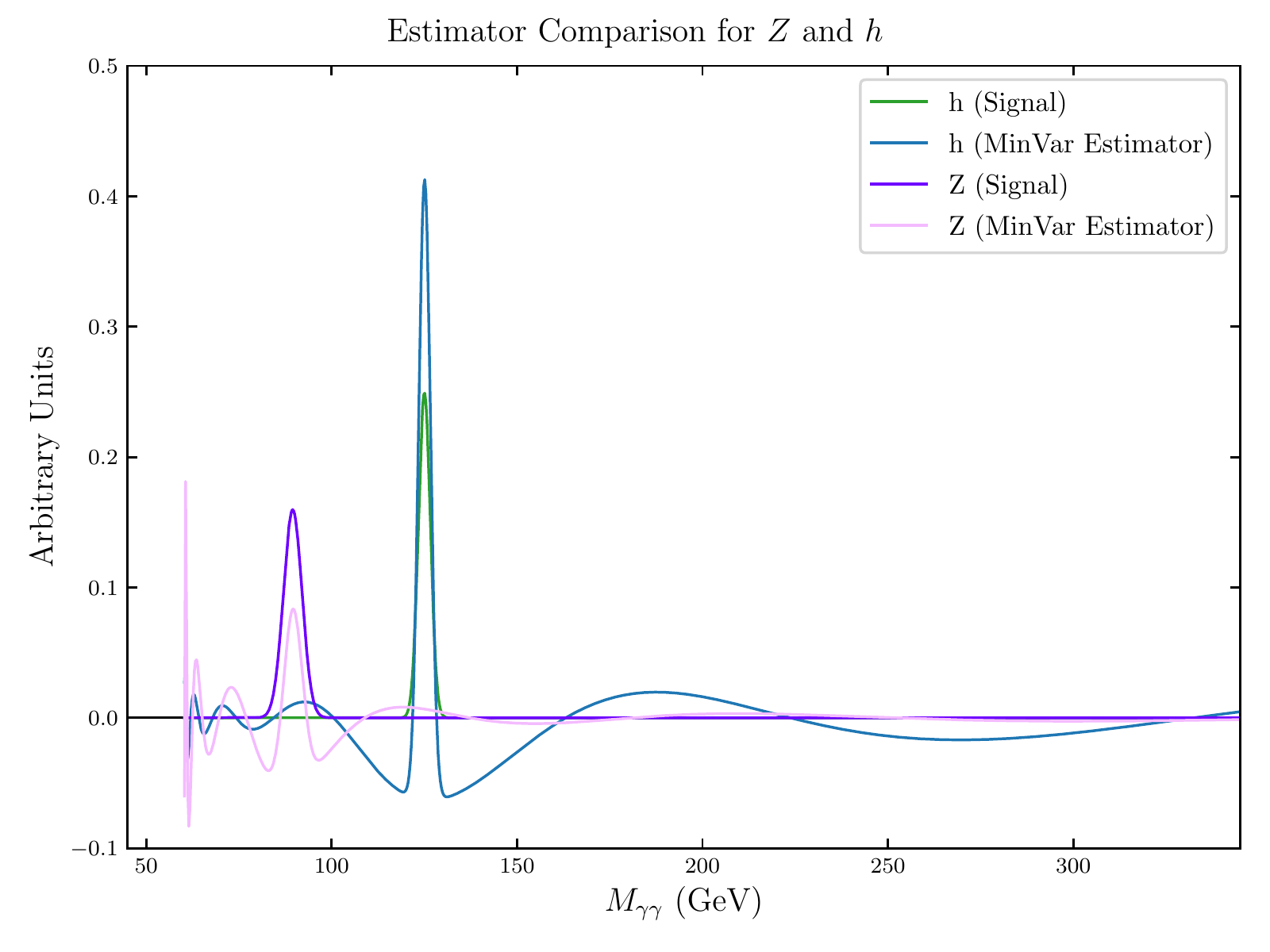}
    \label{fig:est_estimators}
  }

  \subfloat[A comparison between the moments of the test sample (green) and the moments of the minimum-variance estimators for the $Z$ (purple) and $h$ (blue).]{
    \includegraphics[width=0.75\textwidth, clip]{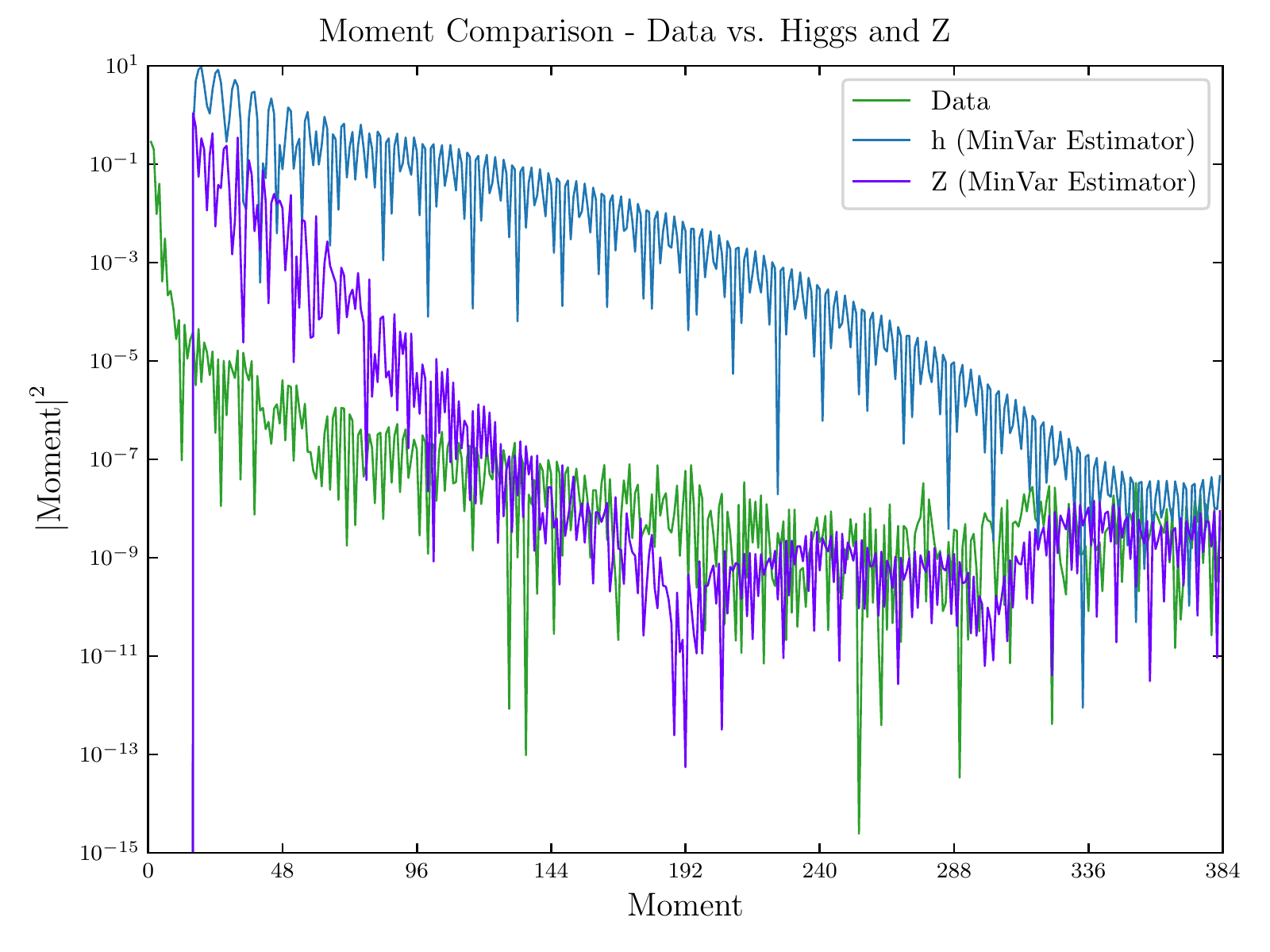}
    \label{fig:est_moments}
  }
  \caption{\label{fig:est}}
\end{center}
\end{figure*}

Because the first \Nmom moments are all free parameters for the background estimation, sensitivity to the resonant contributions lies exclusively in the higher moments (\vecl{f}{n} with $n\geq\Nmom$).  We use minimum-entropy estimators, described in \ref{sec:sigest}, to extract the signal contributions $s_m$:
\begin{align}
  \vecl{\epsilon}{\I ni}{} &= \oplu{\Sigma}{\vec{l}ij}{-1} \veclu{S}{\I n}{j}  \label{eq:minvar_est}\\
  \eta_\I{nm}^{-1}         &= \inner{\epsilon_{\I n}}{S_{\I m}} = \veclu{S}{\I n}{i}\oplu{\Sigma}{ij}{-1}\veclu{S}{\I m}{j} \label{eq:minvar_overlap} \\
  s_\I{n}                  &= \eta_\I{n~}^{\I{~k}} \inner{\epsilon_{\I k}}{f} \label{eq:minvar_full} \, .
\end{align}
Note that the Hilbert space indices in Eq.~\ref{eq:minvar_est} and~\ref{eq:minvar_overlap} are taken to run over the higher moments \textit{only}, that is, $i,j\in\left[\Nmom,\infty\right)$.  This is an important departure from the convention used in the rest of this paper.

Equation~\ref{eq:minvar_est} defines \vecl{\epsilon}{\I n}, the minimum-variance estimator associated with the $n$'th resonant signal.  These estimators, by their definition, are orthogonal to the first \Nmom moments (the background).  However, they are not necessarily orthogonal to one another.  To properly account for multiple signals, the overlap matrix $\eta_{nm}^{-1}$ is introduced, whose inverse orthogonalizes the estimators with respect to the other signals (and also correctly normalizes the estimators).  This procedure is optimal for multiple signals in the sense that it minimizes the uncertainty on each signal contribution individually, as well as the overall uncertainty.

Here the inner products and the matrix multiplications are taken with respect to the higher moments \textit{only}. The covariance $\opl{\Sigma}{\vec l}$, on the other hand, is defined from the \textit{lower} moments:
\begin{equation}
  \opl{\Sigma}{\vec{l}jk} = \sum\limits_{i=1}^{\Nmom-1}\vecu{f}{i}\opl{I}{ijk} \, .
\end{equation}
The estimators retain the properties of linearity and unbiasedness regardless of the choice of \op{\Sigma}.  The choice, then, is motivated by convenience and optimality.  Were the true distribution and its associated covariance $\opu{\Sigma}{\mathrm{T}}$ known \textit{a priori}, one would choose $\op{\Sigma}=\opu{\Sigma}{\mathrm{T}}$ and obtain an optimal set of estimators.  However, given that the true distribution is not known, the next-best choice is to use the empirical covariance obtained from the data.  This is a good approximation, as the covariance matrix is dominantly a function of the lower moments.

Returning to the $M_{\gamma\gamma}$ test spectrum, consider the estimators for the $Z$ and $h$.
A comparison between the signals and their corresponding estimators can be seen in Figure.~\ref{fig:est_estimators}.  The moments of the estimators are shown overlaid with the decomposition of the test data in Figure.~\ref{fig:est_moments}.  The tail visible in the data moments (green curve) from $n=15$ and above is the signal contribution to the data.  Usually this tail would not be visible, but the scaling of the test data reduces the statistical fluctuations and makes it more obvious to the eye.

Once the signal normalizations have been extracted, the background parameters are obtained by subtracting the signal contribution from the data lower moments:
\begin{equation}
  c_n = \left\{\begin{array}{lr}
    \vecl{f}{n} - s^\I{m} \vecl{S}{\I mn}, & n<\Nmom \\
    0                                      & n\geq\Nmom
  \end{array}\right. \, .
\end{equation}
The parameters for the test dataset, both signal and background, are shown in Table~\ref{tab:testdata_model}.

\begin{table}[tb]
\begin{center}
\scriptsize
\def\arraystretch{1.35}
  \begin{tabular}{rS[table-format=2.2e1]rS[table-format=2.2e1]}
\hline\hline
  Parameter & \multicolumn{1}{r}{Value} & Parameter    & \multicolumn{1}{r}{Value} \\
  \hline
     $c_1$  &  5.28e-1                  &     $c_8$    &  1.31e-3  \\
     $c_2$  & -4.41e-1                  &     $c_9   $ & -5.79e-3  \\
     $c_3$  & -9.76e-2                  &     $c_{10}$ & -3.39e-3  \\
     $c_4$  &  1.94e-1                  &     $c_{11}$ &  2.40e-3  \\
     $c_5$  & -1.76e-2                  &     $c_{12}$ &  5.09e-4  \\
     $c_6$  & -5.06e-2                  &     $c_{13}$ & -1.06e-3  \\
     $c_7$  &  1.05e-2                  &     $c_{14}$ &  7.70e-5  \\
  \hline
     $s_Z$  &  9.98e-3                  &     $s_h$    &  2.01e-3  \\
  \hline
  \hline
\end{tabular}
  \caption{The parameters for the FD model of the test dataset. These are computed with $\lambda=32.90~\GeV$, $\alpha=0.60$ and $\Nmom=15$.\label{tab:testdata_model}}
\end{center}
\end{table}

\begin{table}[tb]
\begin{center}
\scriptsize
\def\arraystretch{1.35}
  \begin{tabular}{rS[table-format=5]S[table-format=5(3)]}
\hline\hline
             & {Injected} & {Extracted} \\
  \hline
     $N_h$   &       4000 &   4069(331) \\
     $N_Z$   &      20000 &  20196(933) \\
  \hline
  Corr$_{Zh}$ && \multicolumn{1}{r}{-0.068} \\
  \hline
\end{tabular}
  \caption{ Comparison between the injected and extracted signal for the test dataset. For both the $Z$ and $h$, the difference is well below $1\sigma$ and consistent with the statistical variation of the test dataset.  The correlation between the two extracted signals is small and negative.  \label{tab:testdata_lin}}
\end{center}
\end{table}

The signal parameters are converted to yields by multiplying by the total number of events in the dataset.
The uncertainties can be computed using
\begin{equation}
  \sigma_\I{nm} = M\vecl{\omega}{n} \opl{\Sigma}{\vec f} \vecl{\omega}{m} \label{eq:sigunc} \, ,
\end{equation}
where
\begin{equation}
  \vecl{\omega}{\I ni} = \lambda_\I{n~}^\I{~k} \vecl{\epsilon}{\I ki} \, ,
\end{equation}
essentially the projection of the full covariance matrix onto the subspace defined by the signal estimators.  The extracted signals from the test data, compared to the injected signal strengths, are shown in Table.~\ref{tab:testdata_lin}.  For both the $Z$ and $h$, the difference between injected and extracted signals is small and consistent with the statistical variation of the test dataset ($\pm66$ events for the $h$ and $\pm182$ events for the $Z$).  There is a small negative correlation between the two extracted signals.

  \section{Optimization of Hyperparameters}\label{sec:hyper_opt}
The final ingredient in FD is the selection of the hyperparameters.  For the power-law transformation, there are three hyperparameters: $x_0$, the lower mass limit; $\lambda$, the length scale; and $\alpha$, the scaling exponent.  Additionally there is \Nmom, the number of moments to allocate to background modeling.

The first of these is a true free parameter and serves only to delineate the region of interest.  If the series is not truncated ($\Nmom=\infty$), then $\alpha$ and $\lambda$ are free parameters as well - any value will do, because all choices result in a complete basis.  Unfortunately, this would leave no higher moments with which to search for resonances!

It is clearly desirable to choose the smallest possible \Nmom that allows adequate representation of the smooth background.  Beyond the innate utility of producing an optimal representation of the dataset, this is also a compromise between avoiding loss of signal sensitivity on the one hand (\Nmom too large), and poor modeling with attendant biases on the other (\Nmom too small).  The hyperparameters $\alpha$ and $\lambda$ are chosen to support this end, and to produce the most succinct representation of the data.

  We make the concept of `succinct' concrete using a minimum-description-length (MDL) approach~\cite{9780262072816}.  The objective is to minimize the amount of information required to fully represent the dataset using a two-part encoding scheme.  The expected total information required to encode the dataset in this way is
\begin{equation}
  \mathcal{L} = \Dkl{\vec{f}}{\vec{c} + s^\I{m}\vecl{S}{\I mn} } + \Dkl{\vec{c}}{\vec{p}} \, .\label{eq:mdlKL}
\end{equation}
The first term is the Kullback-Liebler (KL) divergence of the full dataset with respect to the full model, which represents the `information cost' (in nats) to encode the full decomposition \vec{f} if the model parameters are known.  The second term similarly represents the information cost to encode the background estimate \vec{c} given some prior background assumption \vec{p}.  The resonant contributions are assumed to have a constant information cost, independent of hyperparameters, and thus are neglected for the purpose of minimization.

Because the moments are normally distributed, the KL divergences can be calculated using the formula for two multivariate Gaussians:
\begin{equation}
  \Dkl{\vec a}{\vec b} = \frac{1}{2}\left(
             \tr{\oplu{\Sigma}{b}{-1}\opl{\Sigma}{a}}
           + \dt{a}{b}^\top \oplu{\Sigma}{b}{-1} \dt{a}{b}
           - L
           - \ln\det\left(\oplu{\Sigma}{b}{-1}\opl{\Sigma}{a} \right)\right)\, ,
\end{equation}
where \opl{\Sigma}{b} and \opl{\Sigma}{a} are the uncertainty matrices associated with \vec{a} and \vec{b}, respectively, and $L$ is the number of degrees of freedom.

\begin{figure*}[!ptb]
\begin{center}
  \includegraphics[width=\textwidth, clip]{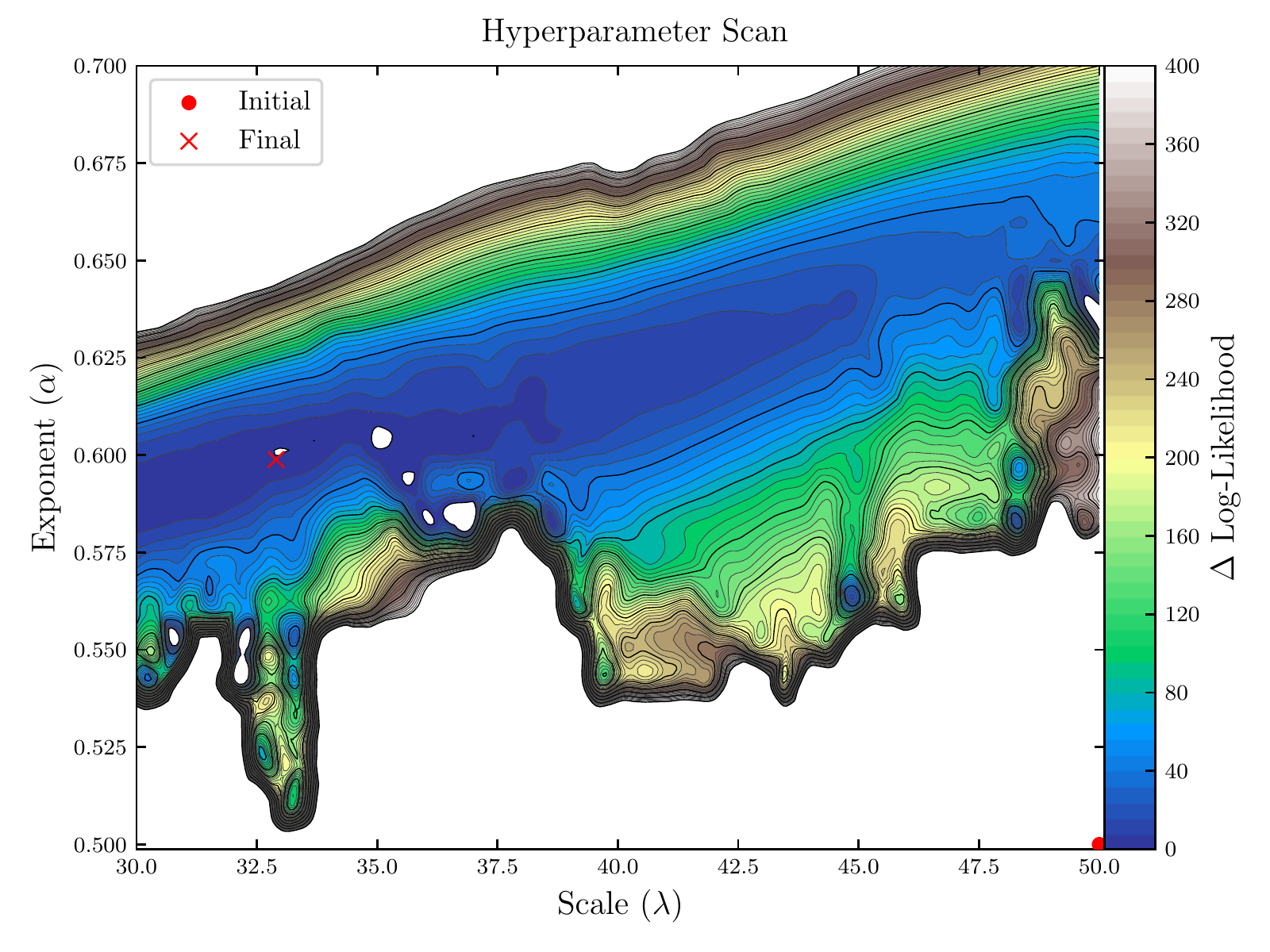}
  \label{fig:hyperparameter_scan}

  \caption{A scan of the hyperparameters $\alpha$ and $\lambda$.  The simulated spectrum has roughly $5\times10^7$ events scaled down to $2\times10^6$ in order to approximate the statistically-asymptotic behavior. The figures shows the difference between the cost function $\mathcal L$ at each combination $\left(\alpha, \lambda\right)$ and the cost function at the minimum $\left(\alpha, \lambda\right)$.  The number of background moments \Nmom is profiled, that is, at each point the \Nmom has been chosen that minimizes $\mathcal L$ at that point.  The red `$\times$' is placed at the minimum, where the hyperparameters are optimal. \label{fig:hyperparameter_scan}}
\end{center}
\end{figure*}

The second term of Eq.~\ref{eq:mdlKL} is amenable to approximation.  If the prior \vec{p} is taken to be weak (\opl{\Sigma}{p} is generally large with respect to \opl{\Sigma}{c}), the first two terms of the Kullback-Liebler divergence approach zero.  The log-determinant approaches $L\ln\frac{M}{j}$, where $j$ is the equivalent statistical strength of the prior (that is, the prior contains equivalent information to $j$ events).  The weakest reasonable prior is $j=L$ (in order for $L$ moments to be independent, they must be based on a distribution of at least $j$ events).  Thus the second term can be approximated
\begin{equation}
  \Dkl{\vec{c}}{\vec{p}} \approx \frac{\Nmom}{2}\ln\left(\frac{M}{\Nmom e}\right) \, .
\end{equation}

To carry out the minimization of Eq.~\ref{eq:mdlKL} requires some care.  Multiple minima are a ubiquitous feature, as might be anticipated by the fact that all $\left(\alpha, \lambda\right)$ produce an exact representation as $\Nmom\rightarrow\infty$.  We find that a two-stage process is most effective: a grid search over a defined range followed by a gradient-descent minimization started from the best point identified in the grid search.  This process typically entails evaluating $\mathcal L$ at several hundred combinations of $\left(\alpha, \lambda\right)$, at a minimum.

If \vec{f} is known, the signal and background contributions $s^\I{m}$ and \vec{c} may be calculated with little computational effort.  However, calculating \vec{f} from a dataset is much more computationally intensive.  Furthermore, the complexity scales linearly with the size of the dataset.  This speed of this procedure can be substantially improved by making use of the transformation matrices defined in Sec.~\ref{sec:xfrm}.

This is accomplished by decomposing the dataset at some initial choice of hyperparameters, $\left(\alpha_\mathrm{ini}, \lambda_\mathrm{ini}\right)$, to derive an initial decomposition \vecl{f}{\mathrm{ini}}.  The appropriate transformation matrices are then applied to extract the decomposition at each point required for the search.  In the test dataset, having $5\times10^7$ events, a transformation is roughly two orders of magnitude faster than a full decomposition.

The results of a scan over the test dataset are shown in Fig.~\ref{fig:hyperparameter_scan}.  From an initial decomposition at $\lambda=45~\GeV$ and $\alpha=0.50$, the scan identifies $\lambda=32.90~\GeV$ and $\alpha=0.60$ as optimal, with $\Nmom=15$.  Several minima are evident.  The `valley' structure arises from the fact that the hyperparameters most strongly influence the decomposition's tail (for finitely many moments, the decomposition always goes like $\simeq e^{-z}$ for sufficiently large $z$).  Outside the valley, the tail is poorly described by $e^{-z}$, requiring larger \Nmom and consequentially producing much more costly decompositions.

\subsection{Positive-definiteness}
This procedure for selecting the hyperparameters also addresses one of the major shortcomings of conventional orthogonal density estimation:  the problem of positive-definiteness.  The underlying probability distribution $f\ofz$, being a probability distribution, must be everywhere non-negative.  However, the Hilbert space of functions that can be represented in the orthonormal basis is more general, and includes functions that take negative values.  Between the statistical uncertainty of a finite dataset and the truncation of the series, there is no guarantee that the resulting approximation of $f\ofz$ will be non-negative, even though $f\ofz$ itself must be.

It turns out that the covariance matrix corresponding to a given model is invertible if and only that model has no zeros.  Because the computation of $\mathcal L$ requires the inversion of the model's covariance matrix, it is defined only when the model has no zeros.  If the model does have zeros, the computation will fail and $\mathcal L$ is assigned an infinite value.  The hyperparameter selection algorithm thus naturally excludes models that are not positive-definite.

\section{Statistical interpretation}\label{sec:statistics}
When one of the signals \vecl{S}{\I m} represents a hypothetical resonance, a precise statistical interpretation of its observed coefficient $s_\I{m}$ is required.  This section describes a convenient approximation for the probability of observing a particular value of $s_\I{m}$ given a model (which may or may not include an actual signal contribution), describes the calculation of p-values and limits, and finally demonstrates these procedures on the test data spectrum.

\subsection{Approximate probability distributions for the signal and background parameters}
The data decomposition \vec{f}, the background parameters \vec{c}, and the signal normalizations $s^\I{m}$ are all described exactly by compound Poission distributions.  To a high degree of accuracy, these can be approximated by multivariate normal distributions having covariances as defined above.  In most cases, this approximation is extremely good.  One exception is the important case of a small signal on the mass distribution's tail.  Here, a better approximation is useful in order to obtain the most accurate confidence intervals and p-values.

This can be framed more precisely.  Given some model
\begin{equation}
  \vecl{\Omega}{i} = \vecl{c}{i} + s^\I{m}\vecl{S}{\I mi}\, ,
\end{equation}
what is the probability distribution $\mathcal P\left(x \middle| \vec{\Omega}\right)$,
where $x = \veclu{\omega}{\I n}{~~i} \vecl{f}{i}$ is the estimate for the parameter $s_{\I n}$?  Like any probability distribution, $\mathcal P$ can be specified exactly in terms of its central moments:
\begin{equation}
  \mu_i\left(\vec{\Omega}\right) = E\left[ \left( x - s_{\I n} \right)^i \right] = \intdz \Omega\ofz \left( \omega_n\ofz - s_{\I n} \right)^i = \vec{\Omega}\left( \opl{\omega}{\I n} -s_{\I n} \right)^i \vec{1} \, .
\end{equation}
where
\begin{equation}
  \opl{\omega}{\I nij} = \intdz \omega_\I{n}\ofz E_i\ofz E_j\ofz = \veclu{\omega}{\I n}{~~k}I_{ijk}
\end{equation}
is the operator representation of $\omega_\I{n}\ofz$.
Note that `central moments' is used as in the statistical literature, and is distinct from the meaning of `moment' otherwise employed in this paper.
We approximate $\mathcal P$ as a shifted, continuous Poisson distribution whose mean, variance and skewness are matched to the first three central moments:
\begin{align}
  a                                          &= M\left.\mu_2^3\left(\vec\Omega\right) \middle/ \mu_3^2\left(\vec\Omega\right) \right. \\
  b                                          &= M\left.\mu_2\left(\vec\Omega\right) \middle/ \mu_3\left(\vec\Omega\right)\right. \\
  k\ofx                                      &= a + b\left(x - s_\I{n} \right) + 0.5 \\
  \mathcal P\left(x\middle|\vec\Omega\right) &= \frac{a^{k\ofx-1}e^{-a}}{\Gamma\left(k\ofx\vphantom{\hat T}\right)} \label{eq:CPD}\, .
\end{align}
This reduces to the normal distribution and to the classical discrete Poisson distribution in the appropriate limits.  It provides an excellent approximation to the exact distribution in all the cases that the authors have examined.

\subsection{P-values and limits}
The results of a search are customarily expressed in the form of p-values and limits.  Both can be obtained from Eq.~\ref{eq:CPD} with $\vec\Omega=\vecl{\Omega}{b}$, that is, using the null hypothesis as model.  In this section, the $n$'th signal is presumed to be the contribution of interest.  The remaining resonant contributions are considered background. The null-hypothesis model \vecl{\Omega}{b} is obtained by completely excluding the $n$'th signal from consideration (the overlap matrix is constructed using only the $n-1$ background resonances and $s_{\I n}$ is set to zero).

The p-value is defined as the probability of obtaining an estimate $x^\prime$ that is as large or larger than actually observed.  Using Eq.~\ref{eq:CPD}, this is given by
\begin{equation}
  P\left(x^\prime > x \right) = \int\limits_x^\infty dx \mathcal{P}\left(x|\vecl{\Omega}{b}\right) = \frac{\Gamma\left(k\left(x\right), a\vphantom{\hat T}\right)}{ \Gamma\left(k\left(x\right)\vphantom{\hat T}\right) } \, ,
\end{equation}
where $\Gamma\left(y,\lambda\right)$ is the upper incomplete gamma function.

Limits can be conveniently calculated using a Bayesian approach.   Assuming a uniform prior on positive signals, the 95\% confidence-level upper limit $x_{95}$ is defined as
\begin{equation}
  \int\limits_0^{x_{95}} \mathcal P\left( x\middle|\vecl{\Omega}{b} + s_{\I n}\vecl{S}{\I n} \right) ds_{\I n} = 0.95 \times \int\limits_0^\infty \mathcal P\left( x\middle|\vecl{\Omega}{b} + s_{\I n}\vecl{S}{\I n} \right) ds_{\I n} \, ,
\end{equation}
where $x$ is fixed to the normalization of the $n$'th signal as observed in the data, and the integral runs over the corresponding true value.  This integral must be performed numerically.  In terms of events, the limit can be written
\begin{equation}
  N_{95} = M \times x_{95} \, ,
\end{equation}
where $M$ is the number of data events.   This can be converted to a cross-section as desired.

\subsection{Application to the test data spectrum}
\begin{figure*}[!ptb]
\begin{center}
  \subfloat[Gaussian width $1.25\%$]{
    \includegraphics[width=0.5\textwidth, clip]{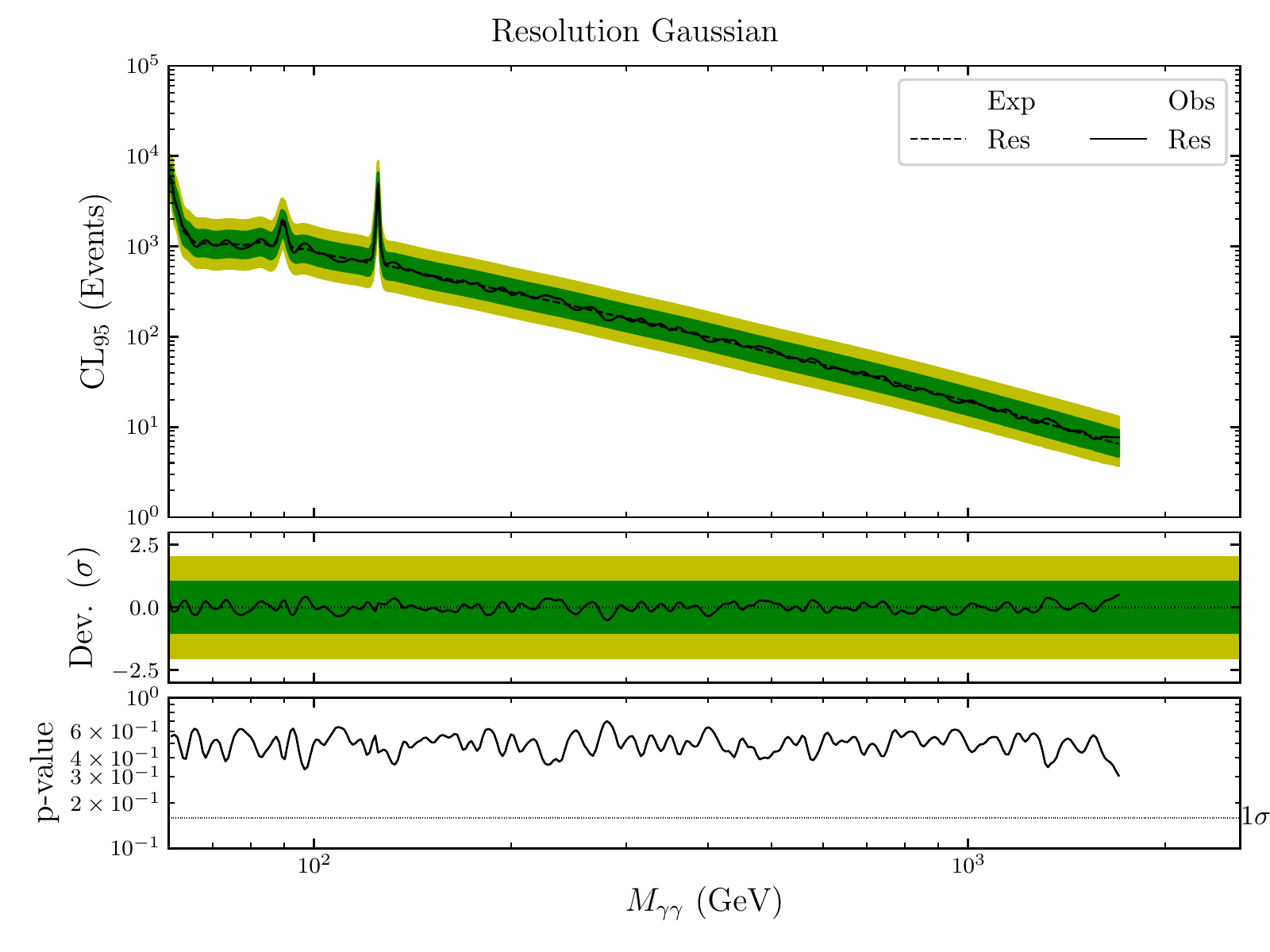}
    \label{fig:scan_gauss_res}
  }
  \subfloat[Gaussian width $2.5\%$]{
    \includegraphics[width=0.5\textwidth, clip]{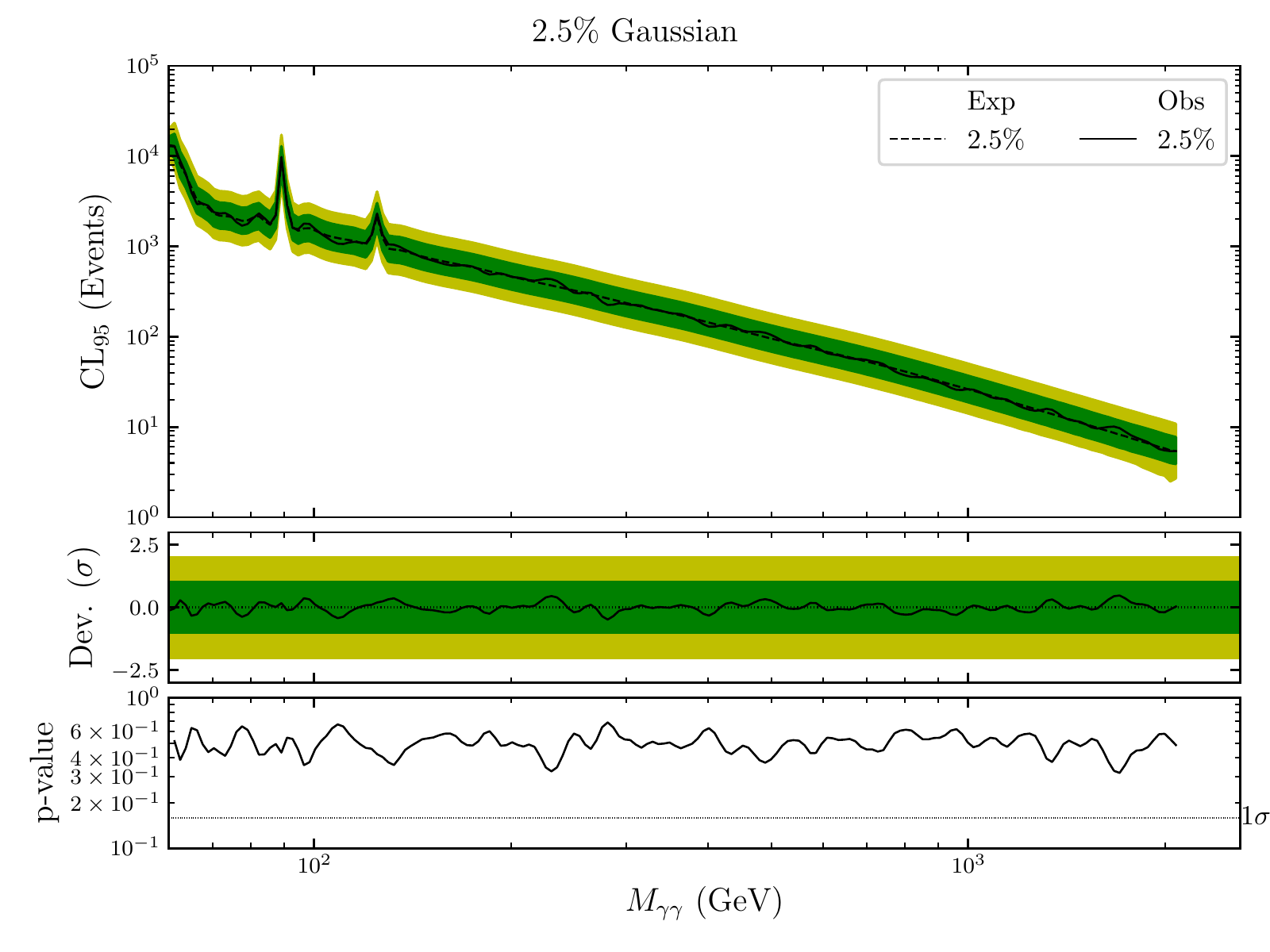}
    \label{fig:scan_gauss_25}
  }

  \subfloat[Gaussian width $5.0\%$]{
    \includegraphics[width=0.5\textwidth, clip]{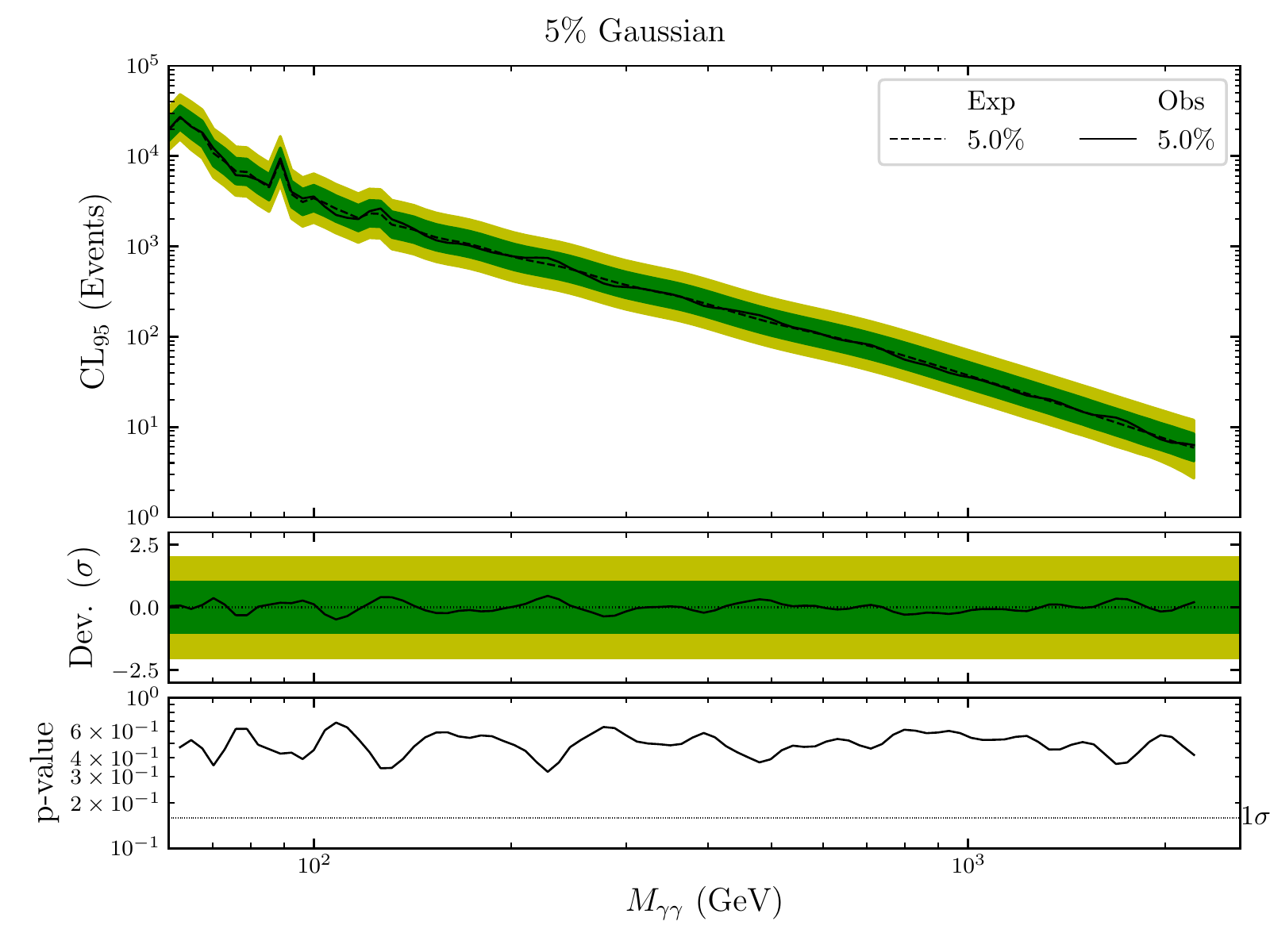}
    \label{fig:scan_gauss_50}
  }
  \subfloat[Gaussian width $10\%$]{
    \includegraphics[width=0.5\textwidth, clip]{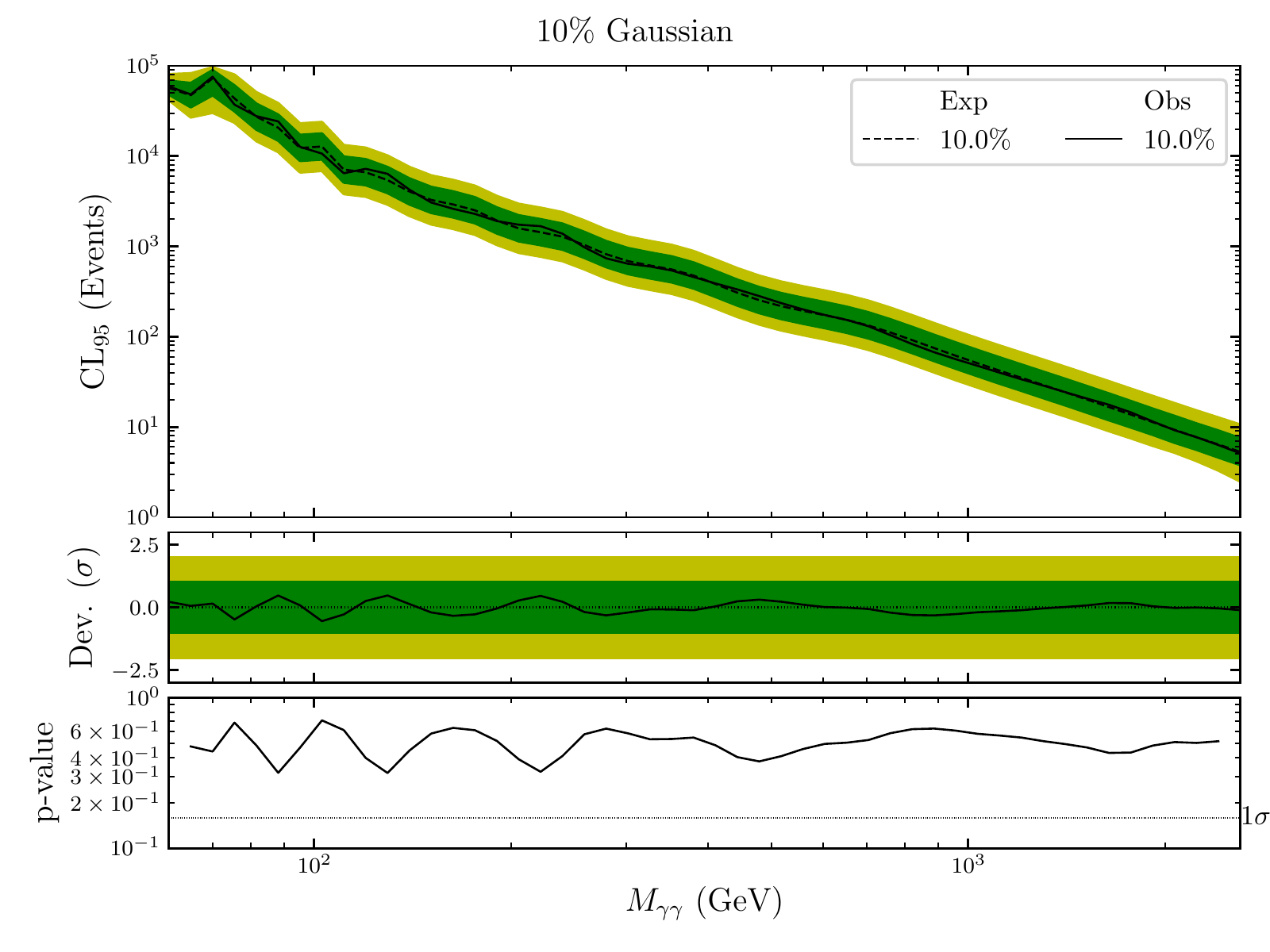}
    \label{fig:scan_gauss_100}
  }

  \caption{Scans over the test-data set.  The test data is scaled down by a factor of $25$, reducing the statistical fluctuations and making the spurious signal (more precisely, the lack thereof) evident.  Also visible is the loss of sensitivity in the vicinity of the $Z$ and $h$, which occurs due to the difficulty of distinguishing between two separate peaks when the peaks have similar masses and widths. \label{fig:scans}}
\end{center}
\end{figure*}

\begin{figure*}[!ptb]
\begin{center}
  \subfloat[Gaussian width $1.25\%$]{
    \includegraphics[width=0.5\textwidth, clip]{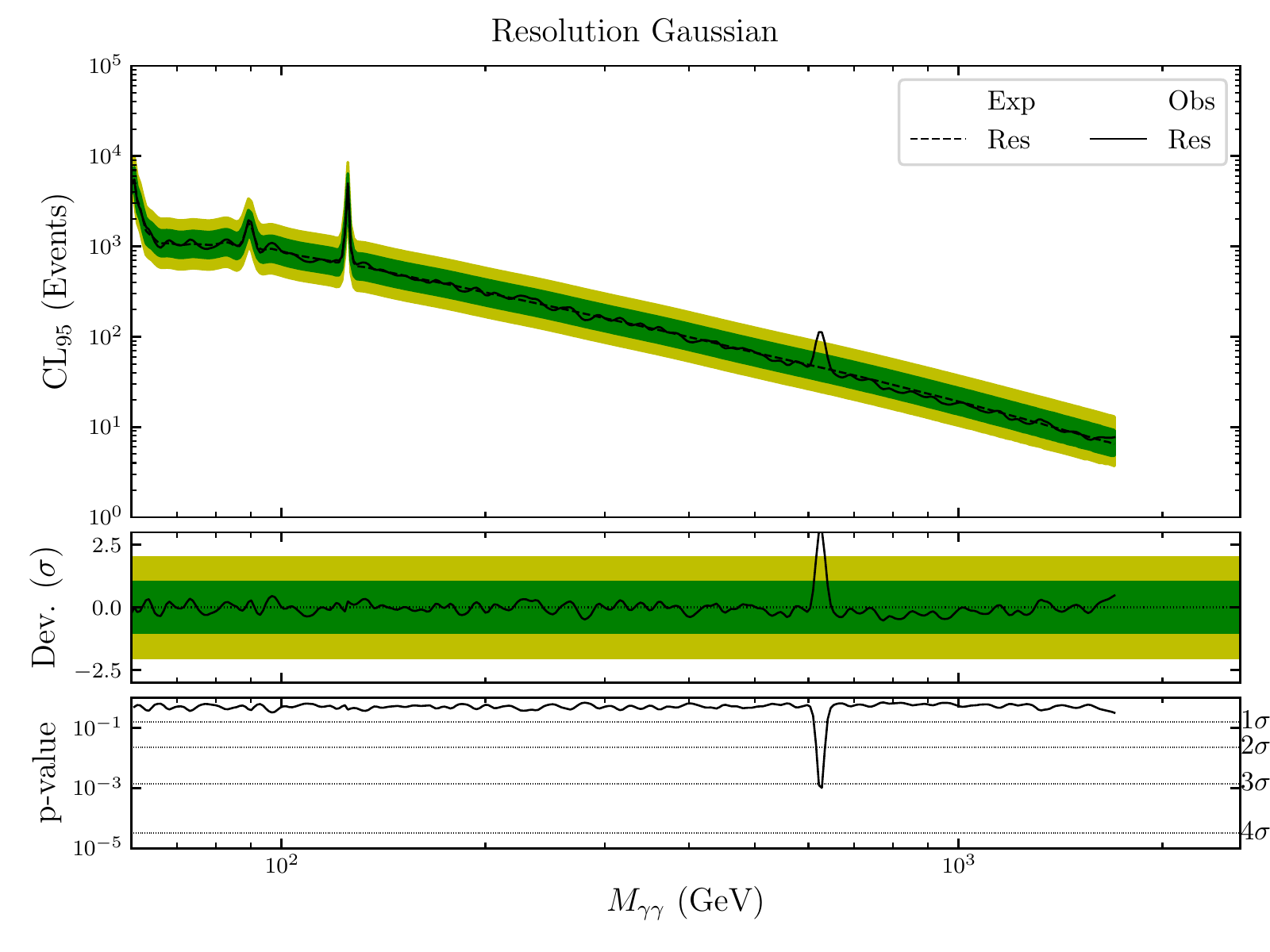}
    \label{fig:scan_inj_gauss_res}
  }
  \subfloat[Gaussian width $2.5\%$]{
    \includegraphics[width=0.5\textwidth, clip]{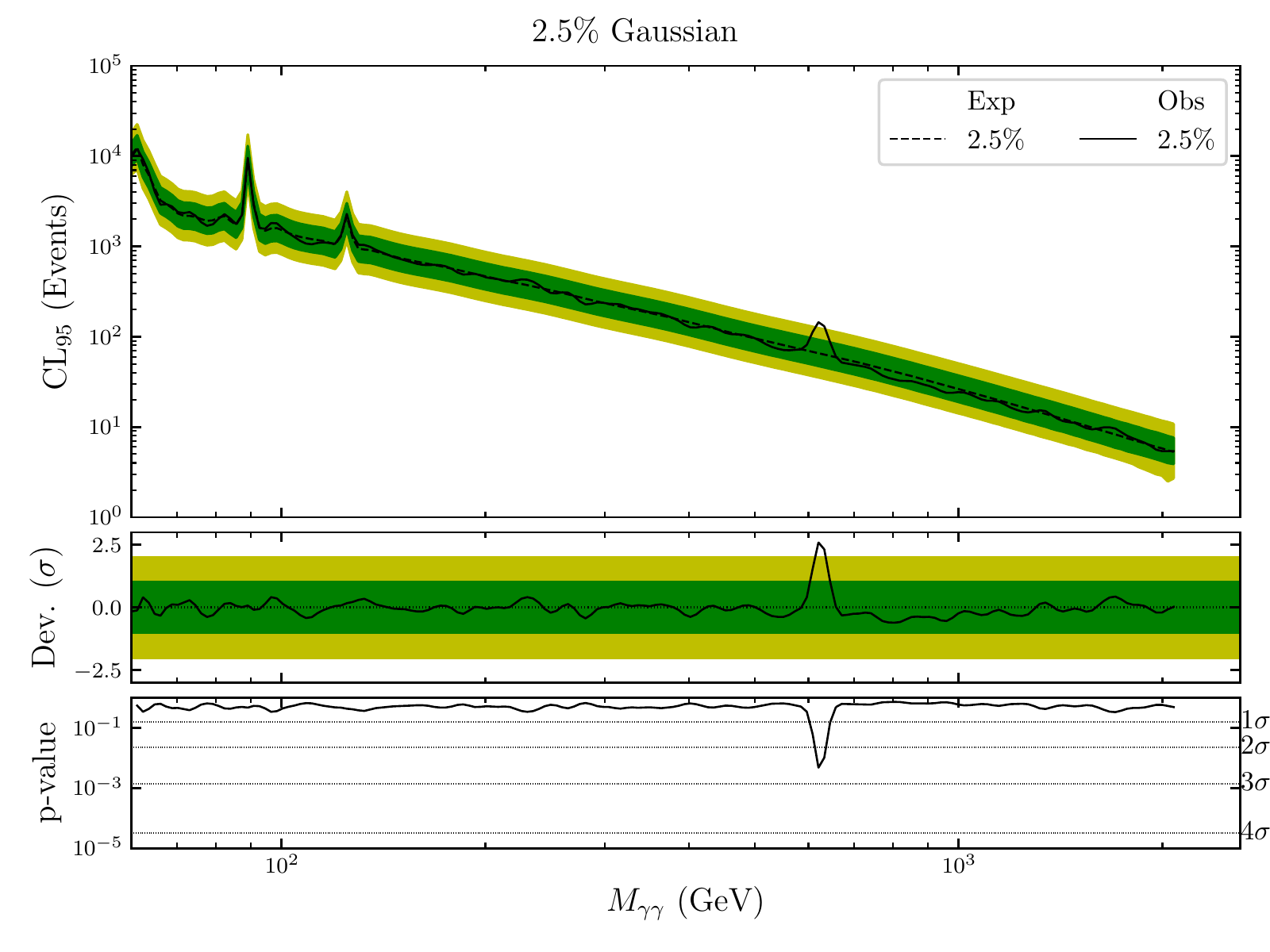}
    \label{fig:scan_inj_gauss_25}
  }

  \subfloat[Gaussian width $5.0\%$]{
    \includegraphics[width=0.5\textwidth, clip]{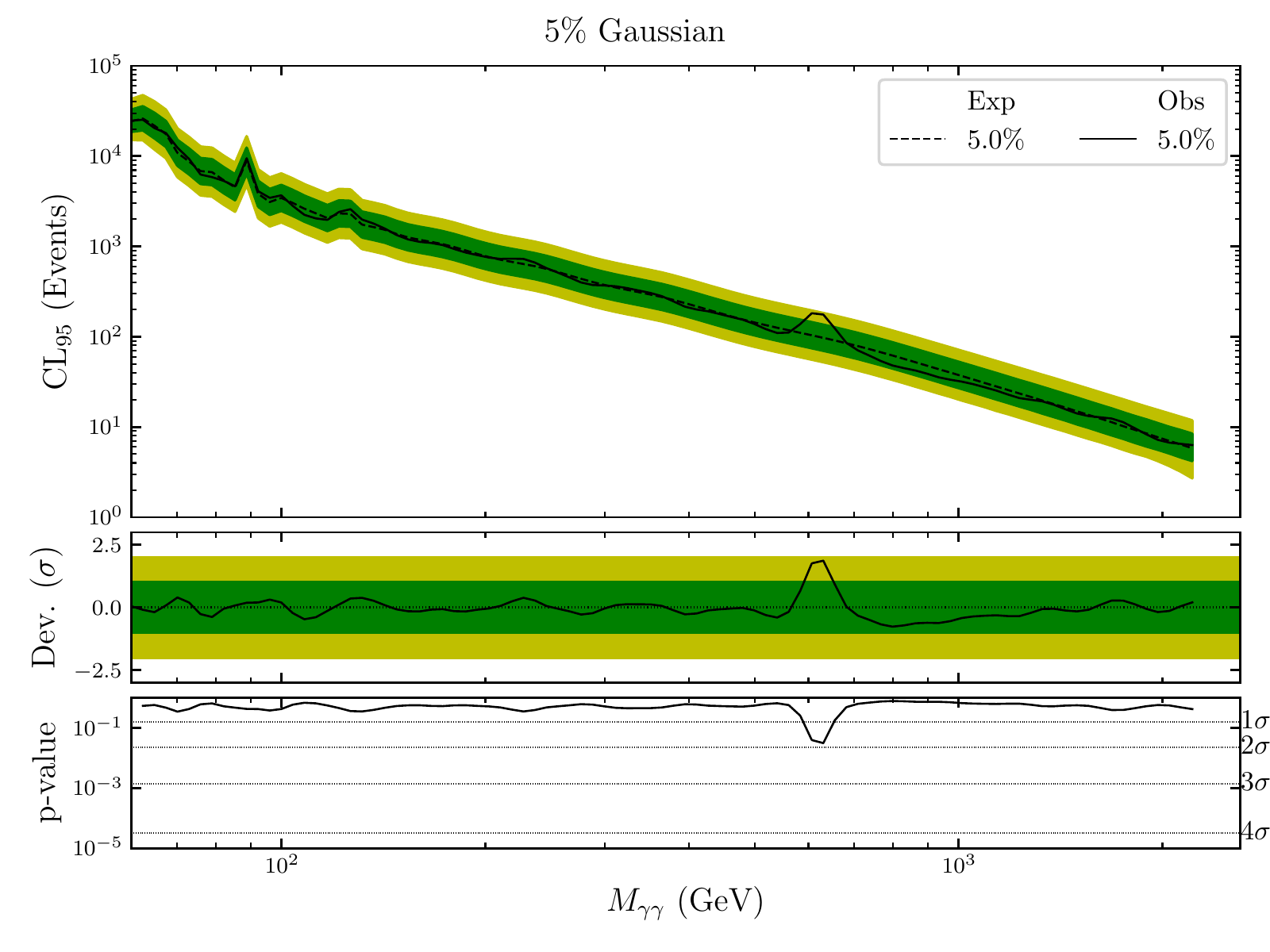}
    \label{fig:scan_inj_gauss_50}
  }
  \subfloat[Gaussian width $10\%$]{
    \includegraphics[width=0.5\textwidth, clip]{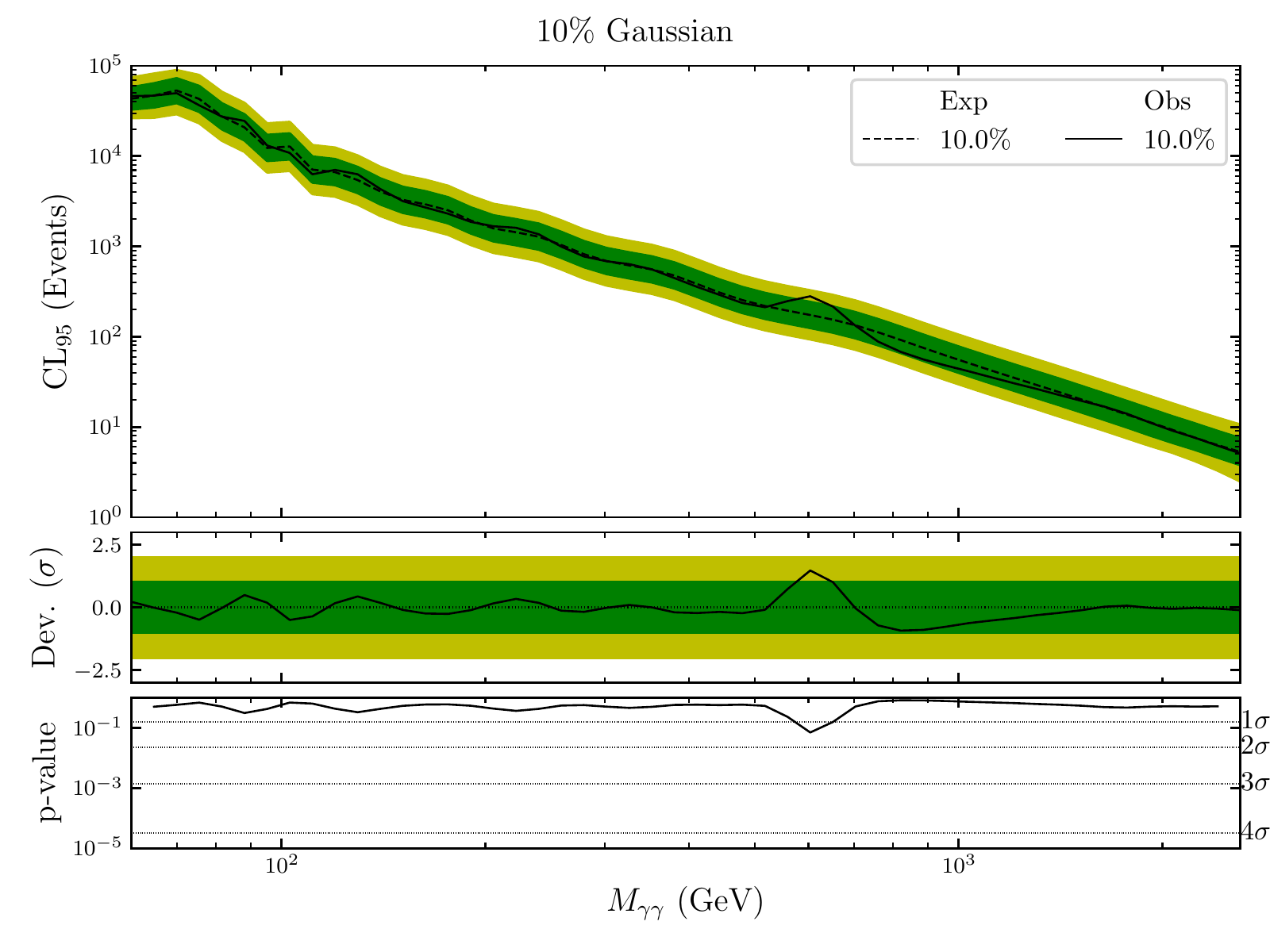}
    \label{fig:scan_inj_gauss_100}
  }

  \caption{Scans over the test-data set where an additional $6.4~\GeV$ wide Gaussian signal has been injected at a mass of $625~\GeV$.  The scans correctly identify the signal, with a p-value of $7.1\times10^{-4}$.  The fit extracts $70\pm22$ events, compared to $66.7$ events injected. \label{fig:scans_inj}}
\end{center}
\end{figure*}

In searches, a data spectrum is typically tested against a collection of hypothetical `new-physics' models, with each model regarded as an independent hypothesis.  For the test dataset, we consider simple Gaussian shapes over a range of masses and widths.  Each model has three resonant contributions: one each for the $Z$ and $h$ (known signals), and one representing a hypothetical unknown resonance.

The p-values and limits on the test spectrum are shown in Figure.~\ref{fig:scans} as a function of the mass and width of the hypothetical resonance.  Because the test spectrum was generated with only the $Z$ and $h$, and no additional resonance, no substantial signal should be detected (the spurious signal should be small).  That is in fact the case, with p-values around 0.5 and small deviations not exceeding $0.5\sigma$.  This indicates minimal bias and spurious signal.

A second set of scans are shown in Figure.~\ref{fig:scans_inj}.  These differ in that an actual third resonance has been injected into the test dataset, a Gaussian with a mass of $625~\GeV$ and width of $6.4~\GeV$, totaling $66.7$ events.  The scans correctly identify the resonance, and assign a p-value of $7.1\times10^{-4}$. The extracted signal is $70\pm22$ events.

An additional interesting feature is visible on the plots - the limits `spike', and  become substantially worse at certain mass/width combinations.  This occurs because of the $Z$ and $h$ contributions, which are treated as free parameters.  If the new-physics signal is too similar to either, the signals become degenerate (or nearly so).  There is consequentially a natural loss in sensitivity to the new-physics signal, as it becomes difficult to distinguish it from the other, known, resonance.

\subsection{Systematic Uncertainties}
Although it is not immediately apparent, the approach presented above correctly and naturally accounts for the systematic uncertainties due to the presence of parameters other than the signal normalization of interest.  This is a result of the fact that each signal estimator is orthogonal to every element of the model other than the corresponding signal.

To see that this is true, consider the model
\begin{equation}
  \vecl{\Omega}{i} = \vecl{c}{i} + s\vecl{S}{i} \, .
\end{equation}
A single signal \vec{S} is considered without loss of generality.  The log-likelihood of the dataset \vec{f} is given by
\begin{equation}
  \mathcal{L} = \frac{1}{2} \left( \vec{\Omega} - \vec{f} \right) \opu{\Sigma}{-1} \left( \vec{\Omega} - \vec{f} \right) + \frac{1}{2}\log\det\left(2\pi\op{\Sigma}\right) \, .\label{eq:approxLLH}
\end{equation}
The profiled likelihood is obtained by fixing $s$ and choosing \vecl{c}{i} to maximize $\mathcal L$
\begin{equation}
  \frac{d\mathcal{L}}{d\vecl{c}{i}} = \vecl{E}{i}\opu{\Sigma}{-1} \left( \vec{\Omega} - \vec{f} \right) \, ,
\end{equation}
which is at an extremum when $\vecl{\Omega}{i}=\vecl{f}{i}$ ($i<\Nmom$).  This is exactly as arranged by the orthogonal estimation procedure of Sec.~\ref{sec:sb_est}.  An equivalent result can be obtained by marginalization.

In reality, \op{\Sigma} is a function of \vecl{c}{i} and $s$ rather than a constant, and Eq.~\ref{eq:approxLLH} is an approximation of the exact likelihood.  In principal this causes some small differences with respect to an exact procedure. However, these corrections are generally proportional to $1/M^2$ and are negligible for all practical purposes.

\section{Summary and discussion; the FD package}\label{sec:summary}
Functional decomposition provides a complete and self-consistent approach to the problem of detecting a narrow, resonant structure superimposed on a smooth background. By employing a carefully-constructed series of orthonormal functions, it is able to successfully model spectra with sculpting or turn-on effects and generalizes to arbitrarily large datasets.  It addresses numerous shortcomings of traditional Monte Carlo and ad-hoc function-based methods.  The mechanism for choosing the series' truncation point strikes a natural balance between sensitivity and flexibility.

The orthonormal exponentials also have application as a means to parameterize falling spectra.  The same algorithm used to create a background for a resonance search also optimally parameterizes the data spectrum using only a handful of coefficients.  This provides a natural way to encapsulate the spectrum's shape without resorting to ad-hoc functions or to reproducing the raw data.

A user-friendly software package that completely implements all of the described techniques is available at \url{https://github.com/ryan-c-edgar/functional-decomposition}.  It is written in Python using Numpy~\cite{Oliphant:2015:GN:2886196}, Scipy~\cite{scipy}, Matplotlib~\cite{Hunter2007} and Numexpr~\cite{numexpr}.  The software can read ROOT ntuples as well as CSV files.  All variables from the input files are available for use; a text-base configuration file specifies which variables to decompose and can optionally specify cuts on any other variables that are available.  The configuration file also allows parametric signal shapes to be freely defined, and can make use of any Python builtins or Numpy/Scipy functions to this end.  It also contains definitions of scan ranges and output plots.

The implementation is highly optimized for speed and memory usage, and consequentially is able to perform fast, unbinned statistical analysis of very large datasets on modest hardware.  Readers are encouraged to download the code and give it a try!

\section*{Acknowledgements}
This work is supported by the U.S. Department of Energy, Office of Science, under grant DE-SC0007859.

\section*{References}
\bibliographystyle{elsarticle/elsarticle-num}
\bibliography{functional_decomp}

\begin{thebibliography}{10}
\expandafter\ifx\csname url\endcsname\relax
  \def\url#1{\texttt{#1}}\fi
\expandafter\ifx\csname urlprefix\endcsname\relax\def\urlprefix{URL }\fi
\expandafter\ifx\csname href\endcsname\relax
  \def\href#1#2{#2} \def\path#1{#1}\fi

\bibitem{Efromovich2010}
S.~Efromovich, \href{https://doi.org/10.1002/wics.97}{Orthogonal series density
  estimation}, Wiley Interdisciplinary Reviews: Computational Statistics 2~(4)
  (2010) 467--476.
\newblock \href {http://dx.doi.org/10.1002/wics.97}
  {\path{doi:10.1002/wics.97}}.
\newline\urlprefix\url{https://doi.org/10.1002/wics.97}

\bibitem{PhysRevD.55.R5263}
{CDF Collaboration}, Search for new particles decaying to dijets at cdf, Phys.
  Rev. D 55 (1997) R5263--R5268,
  \url{https://link.aps.org/doi/10.1103/PhysRevD.55.R5263}.
\newblock \href {http://dx.doi.org/10.1103/PhysRevD.55.R5263}
  {\path{doi:10.1103/PhysRevD.55.R5263}}.

\bibitem{CDF:2009DijetSearch}
{CDF Collaboration}, {Search for new particles decaying into dijets in
  proton-antiproton collisions at $\sqrt{s} = 1.96$~TeV}, Phys. Rev. D 79
  (2009) 112002.
\newblock \href {http://arxiv.org/abs/0812.4036} {\path{arXiv:0812.4036}},
  \href {http://dx.doi.org/10.1103/PhysRevD.79.112002}
  {\path{doi:10.1103/PhysRevD.79.112002}}.

\bibitem{CMS-PAS-EXO-16-056}
C.~Collaboration, {Searches for dijet resonances in pp collisions at
  $\sqrt{s}=13~\mathrm{TeV}$ using data collected in 2016.}, Tech. Rep.
  CMS-PAS-EXO-16-056, CERN, Geneva, \url{http://cds.cern.ch/record/2256873}
  (2017).

\bibitem{dijet2017}
A.~Collaboration, {Search for new phenomena in dijet events using 37 fb$^{-1}$
  of $pp$ collision data collected at $\sqrt{s}=$13 TeV with the ATLAS
  detector}, Phys. Rev. D96~(5) (2017) 052004.
\newblock \href {http://arxiv.org/abs/1703.09127} {\path{arXiv:1703.09127}},
  \href {http://dx.doi.org/10.1103/PhysRevD.96.052004}
  {\path{doi:10.1103/PhysRevD.96.052004}}.

\bibitem{Kautz1954}
W.~Kautz, \href{https://doi.org/10.1109/tct.1954.1083588}{Transient synthesis
  in the time domain}, Transactions of the {IRE} Professional Group on Circuit
  Theory {CT}-1~(3) (1954) 29--39.
\newblock \href {http://dx.doi.org/10.1109/tct.1954.1083588}
  {\path{doi:10.1109/tct.1954.1083588}}.
\newline\urlprefix\url{https://doi.org/10.1109/tct.1954.1083588}

\bibitem{Ross1964}
D.~C. Ross, \href{https://doi.org/10.1109/tcome.1964.6539335}{Orthonormal
  exponentials}, {IEEE} Transactions on Communication and Electronics 83~(71)
  (1964) 173--176.
\newblock \href {http://dx.doi.org/10.1109/tcome.1964.6539335}
  {\path{doi:10.1109/tcome.1964.6539335}}.
\newline\urlprefix\url{https://doi.org/10.1109/tcome.1964.6539335}

\bibitem{Lai1985}
D.~C. Lai, \href{https://doi.org/10.1016/0378-4754(85)90060-6}{Signal
  processing with orthonormalized exponentials}, Mathematics and Computers in
  Simulation 27~(5-6) (1985) 409--420.
\newblock \href {http://dx.doi.org/10.1016/0378-4754(85)90060-6}
  {\path{doi:10.1016/0378-4754(85)90060-6}}.
\newline\urlprefix\url{https://doi.org/10.1016/0378-4754(85)90060-6}

\bibitem{9780262072816}
P.~D. Grunwald, The Minimum Description Length Principle (Adaptive Computation
  and Machine Learning series), The MIT Press, 2007.

\bibitem{Oliphant:2015:GN:2886196}
T.~E. Oliphant, Guide to NumPy, 2nd Edition, CreateSpace Independent Publishing
  Platform, USA, 2015.

\bibitem{scipy}
E.~Jones, T.~Oliphant, P.~Peterson, et~al.,
  \href{http://www.scipy.org/}{{SciPy}: Open source scientific tools for
  {Python}}, [Online; accessed 10 May 2018] (2001--).
\newline\urlprefix\url{http://www.scipy.org/}

\bibitem{Hunter2007}
J.~D. Hunter, \href{https://doi.org/10.1109/mcse.2007.55}{Matplotlib: A 2d
  graphics environment}, Computing in Science {\&} Engineering 9~(3) (2007)
  90--95.
\newblock \href {http://dx.doi.org/10.1109/mcse.2007.55}
  {\path{doi:10.1109/mcse.2007.55}}.
\newline\urlprefix\url{https://doi.org/10.1109/mcse.2007.55}

\bibitem{numexpr}
D.~Cooke, et~al., \href{https://github.com/pydata/numexpr}{{numexpr}: Fast
  numerical array expression evaluator for {Python}, {NumPy}, {pandas},
  {bcolz}, and more}, [Online; accessed 10 May 2018] (2009--).
\newline\urlprefix\url{https://github.com/pydata/numexpr}

\end{thebibliography}

\appendix

\section{Proofs and Derivations}
\subsection{Completeness of the exponentials}\label{sec:completeness}
This section demonstrates that the set of exponentials as defined in Eq. \ref{eq:nonorthodef} is complete with respect to the set of normalizable probability distributions on $\left[0,\infty\right)$.  This is accomplished by showing completeness for a more general category of functions.

Suppose that $f\ofz$ is a real-valued function defined on $\left[0, \infty\right)$ and furthermore that
\begin{equation}
  \lim\limits_{z\rightarrow\infty} f\ofz = 0 \, .
\end{equation}
Consider the transformation $z = -\ln y$.  This bijectively maps the exponentials $\sqrt{2}F_n=e^{-nz}$ on $\left[0,\infty\right)$ to the polynomials $F_n^\star$ on $\left(0,1\right]$:
\begin{equation}
  F_n^\star\ofy = \sqrt{2}y^n \label{eq:xfrmb}\, ,
\end{equation}
and maps the inner product as
\begin{equation}
  \inner{f^{\star}}{g^{\star}} = \int\limits_0^1\frac{dy}{y} f^\star\ofy g^\star\ofy\, .
\end{equation}
Then by the completeness of the polynomials, the transformed function $f^\star\ofy = f\left(-\ln y\right)$ can be represented
\begin{equation}
  f^\star\ofy = \sum\limits_{n=0}^\infty a_n^\star y^n \, .
\end{equation}
However, $\lim\limits_{y\rightarrow0}f^\star\ofy =0 $ and so the constant term $a_0$ must be zero.  It then follows from Eq.~\ref{eq:xfrmb} and the definition of $F_n$ that
\begin{equation}
  f\ofz = \sum\limits_{n=1}^\infty a_n F_n\ofz \, ,
\end{equation}
where $a_n = a_n^\star/\sqrt{2}$.

\subsection{Coefficients of the orthonormal exponentials and recurrence relations}\label{sec:orthcoeffs}
This section constructs an explicit solution for the coefficients of the orthonormal exponentials in terms of the non-orthonormal exponentials.

Consider the functions
\begin{equation}
  \begin{split}
    \Lambda_n\left(t\right) &= \inner{E_n}{ \sqrt{2}e^{-tz}} \\
                            &= \intdz E_n\ofz \sqrt{2}e^{-tz} \label{eq:lamdef}
  \end{split}
\end{equation}
with respect to a complex argument $t$ having $\repart{t}>0$. When $t$ is a positive integer, $\Lambda_n\left(t\right) = \inner{E_n}{F_{c}}$.  Using Eq.~\ref{eq:ortho_def}, $\Lambda_n$ may be written
\begin{equation}
  \Lambda_n\left(t\right) = \sum\limits_{i=1}^n \frac{2\opl{d}{ni}}{t+i} \label{eq:lamexp}\, .
\end{equation}
This is a rational function of $t$, and may be written as the ratio of two polynomials in $t$.  The denominator is at most degree $n$, and the numerator is at most degree $n-1$.  By construction, it is zero for $t\in\left\{1,\dots,n-1\right\}$.  These $n-1$ zeros then specify the form of the numerator exactly (up to a multiplicative factor that is constant in $t$):
\begin{equation}
  f\left(n\right)\prod\limits _{i=1}^{n-1}\left(t-i\right)\, .
\end{equation}
It follows that $\opl{d}{ni}\neq0~\forall i<n$ (otherwise the degree of the numerator would be less than $n$).  Consequentially, the denominator is the product of the $n$ denominators in Eq.~\ref{eq:lamexp}.  Then $\Lambda_n$ may be written
\begin{equation}
  \Lambda_n\left(t\right) = \frac{f\left(n\right)}{t-n}\prod\limits_{i=1}^n\frac{t-i}{t+i}\label{eq:lamfin} \, .
\end{equation}

Next, note that $\Lambda_n$ can be uniquely analytically continued to the full complex plane, and consider the equality between Eq.~\ref{eq:lamexp} and Eq.~\ref{eq:lamfin}:
\begin{equation}
  \sum\limits_{i=1}^n \frac{2\opl{d}{ni}}{t+i} = \frac{f\left(n\right)}{t-n}\prod\limits_{i=1}^n\frac{t-i}{t+i}\, .
\end{equation}
The values of $\opl{d}{nm}$ can be extracted from the residues of the $n$ simple poles at $m=-n,\dots,1$:
\begin{align}
  \opl{d}{nm} &= \frac{1}{2}\lim\limits_{t\rightarrow-m}\left[
    \left(t+m\right)\frac{f\left(n\right)}{t-n}    \prod\limits_{i=1}^n\frac{t-i}{t+i}
            \right] \\
  \opl{d}{nm} &= f\left(n\right)\frac{m}{n+m}\left(-1\right)^{n+m}
            \prod\limits _{i=1}^{m-1}\frac{m+i}{m-i}
            \prod\limits _{i=m+1}^{n}\frac{i+m}{i-m}\, .
\end{align}
Finally, we fix $f\left(n\right)$ from the requirement that $\innerEE{n}{n}$ is to be unity:
\begin{equation}
  \innerEE{n}{n} = \intdz \sum\limits_{i=1}^{n}\opl{d}{ni}\sum\limits _{j=1}^{n}\opl{d}{nj} e^{-\left(i+j\right)z}
                 = \sum\limits_{i=1}^{n}\opl{d}{ni}\sum\limits _{j=1}^{n}\frac{2\opl{d}{nj}}{i+j}
                 = \sum\limits_{i=1}^{n}\opl{d}{ni}\Lambda_n\left(i\right)\, .
\end{equation}
The sum reduces to a single term because $\Lambda_n\left(i\right)$ is nonzero for integer $i$ only if $i\geq n$, so $\innerEE{n}{n} = \opl{d}{nn}\Lambda_n\left(n\right)=f^2\left(n\right)/4n$.  From this, $f\left(n\right)=2\sqrt{n}$ and
\begin{align}
  \Lambda_n\left(t\right) &= \frac{2\sqrt{n}}{t-n}\prod\limits_{i=1}^n\frac{t-i}{t+i} \\
  \opl{d}{nm} &= \sqrt{n}\left(-1\right)^{n+m} \left(\frac{2m}{n+m}\right)
                   \prod\limits _{i=1}^{m-1}\frac{m+i}{m-i}
                   \prod\limits _{i=m+1}^{n}\frac{i+m}{i-m} \, .
\end{align}
The first few of these coefficients are tabulated in Table~\ref{tab:func_coeffs}.

\begin{table}[tb]
\begin{center}
\scriptsize
\def\arraystretch{1.35}
\begin{tabular}{rrrrrrrrr}
\hline\hline
  $\opl{d}{n1}$& $\opl{d}{n2}$  &  $\opl{d}{n3}$ &  $\opl{d}{n4}$ &  $\opl{d}{n5}$ &  $\opl{d}{n6}$ &  $\opl{d}{n7}$ &  $\opl{d}{n8}$ &  $\opl{d}{n9}$  \\
  \hline
  $    1\sqrt{1}$&                  &                  &                  &                  &                  &                  &                  &                   \\
  $   -2\sqrt{2}$& $      3\sqrt{2}$&                  &                  &                  &                  &                  &                  &                   \\
  $    3\sqrt{3}$& $    -12\sqrt{3}$& $     10\sqrt{3}$&                  &                  &                  &                  &                  &                   \\
  $   -4\sqrt{4}$& $     30\sqrt{4}$& $    -60\sqrt{4}$& $     35\sqrt{4}$&                  &                  &                  &                  &                   \\
  $    5\sqrt{5}$& $    -60\sqrt{5}$& $    210\sqrt{5}$& $   -280\sqrt{5}$& $    126\sqrt{5}$&                  &                  &                  &                   \\
  $   -6\sqrt{6}$& $    105\sqrt{6}$& $   -560\sqrt{6}$& $   1260\sqrt{6}$& $  -1260\sqrt{6}$& $    462\sqrt{6}$&                  &                  &                   \\
  $    7\sqrt{7}$& $   -168\sqrt{7}$& $   1260\sqrt{7}$& $  -4200\sqrt{7}$& $   6930\sqrt{7}$& $  -5544\sqrt{7}$& $   1716\sqrt{7}$&                  &                   \\
  $   -8\sqrt{8}$& $    252\sqrt{8}$& $  -2520\sqrt{8}$& $  11550\sqrt{8}$& $ -27720\sqrt{8}$& $  36036\sqrt{8}$& $ -24024\sqrt{8}$& $   6435\sqrt{8}$&                   \\
  $    9\sqrt{9}$& $   -360\sqrt{9}$& $   4620\sqrt{9}$& $ -27720\sqrt{9}$& $  90090\sqrt{9}$& $-168168\sqrt{9}$& $ 180180\sqrt{9}$& $-102960\sqrt{9}$& $  24310\sqrt{9}$ \\
  \hline
  \hline
\end{tabular}
  \caption{The coefficients of the first few orthonormal exponentials.  The $n$'th orthonormal exponential is written: $E_n\ofz=\sqrt{2}\sum\limits_{i=0}^{n} \opl{d}{ni} e^{-iz}. $ \label{tab:func_coeffs}}
\end{center}
\end{table}

Two recurrence relations derived from this result are also useful. The first arises from considering the ratio
$\opl{d}{n\left(m+1\right)}/\opl{d}{nm}$.  This results in a recurrence relation on the coefficients themselves, which is given by
\begin{align}
\begin{split}
  \opl{d}{n1}                &=\left(-1\right)^{n+1}n\sqrt{n}              \\
  \opl{d}{n\left(m+1\right)} &=\frac{m^{2}-n^{2}}{m\left(m+1\right)} \opl{d}{nm}  \label{eq:dn0}\, .
\end{split}
\end{align}
This form can be used to conveniently generate the coefficients for the $n$'th orthonormal exponential with minimal computational effort.  The second is a three-term recurrence relation on the formalized exponentials.  That such a recurrence relation exists is implied by the isomorphism between the orthonormal exponentials and the polynomials (\ref{sec:completeness}).  It is given by:
\begin{align}
\begin{split}
  \Eb{1}\ofz   &= \sqrt{2} e^{-z} \\
  \Eb{n+1}\ofz &= \frac{1}{\phi_{2n+1}}\left(
                               4e^{-z}\Eb{n}\ofz 
                             - \frac{2}{\phi^2_{2n}}\Eb{n}\ofz 
                             - \phi_{2n-1} \Eb{n-1}\ofz
                  \right) \\
  \phi_n       &= \sqrt{1-\frac{1}{n^2}} \label{eq:recurrence} \, .
\end{split}
\end{align}
That Eqs.~\ref{eq:dn0} satisfy this relation can be shown using only simple (though tedious) algebra.  Equation~\ref{eq:recurrence} is generally the fastest and most numerically stable method to evaluate the orthonormal exponentials. 

\subsection{The general hyperparameter transformation matrix}\label{sec:xfrm_general}
This section demonstrates a general result that, under certain conditions, the transformation matrix between two different sets of hyperparameters $\theta$ and $\theta^\star$ can be expressed as a matrix exponential.  This can be seen more generally by noting that the hyperparameter transformations form a Lie algebra.  This section is included nonetheless, first for completeness but also so as to have the result expressed in the most convenient form for the needs of this paper.

Suppose some function $f\ofx$ has a known decomposition in a basis defined by transformed variable $z^\star=T\left(x, \theta^\star\right)$, and the decomposition is desired with respect to a different choice of hyperparameters, $z=\left(x, \theta\right)$.  In the starred basis, the known decomposition is expressed
\begin{equation}
  f\ofx=\vecu{f}{^\star n} E_n\ofzs \, .
\end{equation}
The decomposition in $z$ is related by a linear transformation:
\begin{align}
  \vecl{f}{n} &= \Mf\vecu{f}{\star m} \\
  \Mf         &= \intdz E_n\ofz E_m\ofzs \, .
\end{align}

Now suppose that the hyperparameters are a function of some variable $\beta$, that is, $\theta=\Theta\left(\beta\right)$ with $\Theta\left(0\right)=\theta^\star$.  If $\Theta$ is differentiable with respect to $\beta$, then the transformation matrix is also differentiable with respect to $\beta$.  Its derivative is
\begin{align}
  \frac{d\Mf}{d\beta} &= \intdz E_n^\prime\ofz E_m\ofzs \frac{dz}{d\beta} \\
                      &= \Mfi \intdz E_n^\prime\ofz E_i\ofz \frac{dz}{d\beta} \, .
\end{align}
To obtain the second equation, the term $E_m\ofzs$ has been transformed into a function of $z$ using \op{\mathcal M}.  This results in a matrix differential equation, and if $\frac{dz}{d\beta}$ is a constant, this has the solution
\begin{equation}
\begin{split}
  \op {\mathcal M}     &= \exp\left[ \op{\mathcal{T}} \right] \\
  \opl{\mathcal T}{nm} &= \intdz E_n^\prime\ofz E_m\ofz \frac{dz}{d\beta} \, , \label{eq:HPxfrm}
\end{split}
\end{equation}
where  $\exp$ is the matrix exponential.

\subsection{Hyperparameter transformation matrix for the power-law transformation}\label{sec:xfrm_powerlaw}
The power-law transformation is written
\begin{equation}
  z = \left(\frac{x-x_0}{\lambda}\right)^\alpha \, .
\end{equation}
Suppose that $\alpha=\alpha\left(\beta\right)$ and $\lambda=\lambda\left(\beta\right)$.  Then
\begin{equation}
  \frac{dz}{d\beta} = \frac{1}{\alpha} \frac{d\alpha}{d\beta} z\ln z-\frac{\alpha}{\lambda}\frac{d\lambda}{d\beta}z \, .
\end{equation}
For this to be a constant, as required for Eq.~\ref{eq:HPxfrm}, both terms must individually be constant:
\begin{align}
  \frac{1}{\alpha} \frac{d\alpha}{d\beta}          &= c \label{eq:consC} \\
  -\frac{\alpha}{\lambda}\frac{d\lambda}{d\beta}   &= s \label{eq:consS}\, .
\end{align}
From Eq.~\ref{eq:consC}, it follows that $\alpha=\alpha^\star e^{\beta c}$.  Then substituting this into Eq.~\ref{eq:consS},
\begin{align}
  \frac{d}{d\beta}\ln\lambda         &= -\frac{s}{\alpha^{\star}}e^{-\beta c} \\
  \ln\lambda+C                       &= \frac{s}{\alpha^{\star}c}e^{-\beta c} \\
  \ln\frac{\lambda}{\lambda^{\star}} &= \frac{s}{\alpha^{\star}c}\left(e^{-\beta c}-1\right)
\end{align}
where the constant C has been fixed by the requirement that $\lambda\left(0\right)=\lambda^\star$.  At $\beta=1$, these yield expressions for the constants $s$ and $c$:
\begin{align}
  c &= \ln\frac{\alpha}{\alpha^\star}                              \label{eq:ExpC} \\
  s &= -\frac{\alpha c}{e^c - 1} \ln\frac{\lambda}{\lambda^\star}  \label{eq:ExpS} \, .
\end{align}

From Eq.~\ref{eq:HPxfrm}, the infinitesimal transformation matrix can be expressed as
\begin{equation}
  \opl{\mathcal T}{nm} = \intdz E_n^\prime\ofz E_m\ofz \left( cz\ln z + sz \right) \, ,
\end{equation}
It is easiest to evaluate the integrals in the exponential basis and then transform to the orthonormal basis.  Using the series coefficients \opl{d}{nm} derived in \ref{sec:orthcoeffs}, the argument to the matrix exponential can be written
\begin{equation}
  \intdz E_n^\prime\ofz E_m\ofz \left( cz\ln z + sz \right) = -\sum\limits_{i=1}^\infty\sum\limits_{j=1}^\infty d_{ni}d_{mj}i\intdz e^{-\left(i+j\right)z} \left( cz\ln z + sz \right) \label{eq:HParg} \, .
\end{equation}
This can be evaluated numerically with the aid of the integrals
\begin{align}
  \intdz e^{-nz}z       &=\frac{1}{n^{2}} \\
  \intdz e^{-nz}z\log z &=\frac{1-\gamma-\log n}{n^{2}} \, ,
\end{align}
where $\gamma$ is the Euler-Mascheroni constant.  This finally gives an expression for the transformation matrix,
\begin{equation}
\begin{split}
  \op {\mathcal{M}} &= \exp\left[ c\op C + s\op S  \right] \\
  \opl{C}{nm}       &= -\sum\limits_{i=1}^n\sum\limits_{j=1}^m \opl{d}{ni}\opl{d}{mj}\frac {i}{\left(i+j\right)^2} \left[ 1-\gamma-\ln\left(i+j\right) \right] \\
  \opl{S}{nm}       &= -\sum\limits_{i=1}^n\sum\limits_{j=1}^m \opl{d}{ni}\opl{d}{mj}\frac {i}{\left(i+j\right)^2} \\
\end{split}
\end{equation}
where $s$ and $c$ are set according to the desired start and end hyperparameter values as per Eq.~\ref{eq:ExpC} and ~\ref{eq:ExpS}.

\subsection{Finiteness of the mean and variance of the moments}\label{sec:meanvar}
We here argue that the empirical moments \vecl{f}{n} corresponding to some function $f\ofx$ must be normally distributed in the high-statistics limit.  That is, the conditions of the central limit theorem always apply, regardless of the underlying function.

Suppose that $f\ofx$ is a continuous probability distribution that is everywhere finite on $\left[x_0,\infty\right)$.
Then the moments and (co-)variance may be expressed
\begin{align}
  \vecl{f}{n}      &= \intdz f\ofz E_n\ofz \\
  \opl{\Sigma}{nm} &= \intdz f\ofz E_n\ofz E_m\ofz - \vecl{f}{n} \vecl{f}{m} \, .
\end{align}
Because the orthonormal exponentials are linear combinations of terms like $\exp\left(-kz\right)$, both of these may be expressed as linear combinations of integrals of the form
\begin{equation}
  M_k = \intdz f\ofz e^{-kz} \,.
\end{equation}
Comparing the integrands for $M_k$ and $M_j$, if $j<k$ then
\begin{equation}
  \left| f\ofz \right| e^{-kz} < \left| f\ofz \right| e^{-jz} \, .
\end{equation}
The convergence of the longest length-scale integral (i.e., \vecl{f}{1}) then ensures the convergence of all the smaller length-scale integrals.  A necessary and sufficient condition for the moments (and covariance) to be finite is then
\begin{equation}
  \intdz f\ofz e^{-z} \in \R \, \label{eq:momcond}.
\end{equation}
The previously-derived condition for completeness, $\lim\limits_{z\rightarrow\infty} f\ofz = 0$, is more restrictive. Thus the moments, their variance, and the covariance between any two moments must all be finite.

\subsection{Calculating the covariance matrix}\label{sec:funcovfact}
We here record a surprising (and convenient) result: the empirical covariance matrix need not be evaluated directly, because the $N\times N$ covariance matrix can be calculated from the first $2N$ moments.  This is most readily seen from the continuous analogue of Eq.~\ref{eq:empcov}:
\begin{equation}
  \opl{\Sigma}{nm} = \intdz f\ofz E_n\ofz E_m\ofz - \vecl{f}{n}\vecl{f}{m} \label{eq:covdef}\, .
\end{equation}
Substituting $f\ofz$ with its expansion,
\begin{equation}
  \opl{\Sigma}{nm} = \intdz\vecu{f}{i} E_i\ofz E_n\ofz E_m\ofz - \vecl{f}{n}\vecl{f}{m} \, .
\end{equation}
Note that the triple-integral is nonzero only if $\left|n-m\right|\leq i\leq n+m$.  This is because $E_n E_m$ is contained in the subspace spanned by $\left\{E_0,\dots,E_{n+m}\right\}$, to which $E_i$ is by definition orthogonal if $i>n+m$.  The rest of the condition follows by permutation of the indices.  The covariance matrix can therefore be represented
\begin{equation}
\begin{split}
  \opl{\Sigma}{nm} &= \sum\limits_{i=\left|n-m\right|}^{n+m} \vecu{f}{i}\opl{I}{inm} - \vecl{f}{n}\vecl{f}{m} \\
  \opl{I}{ijk}     &= \intdz E_i\ofz E_j\ofz E_j\ofz \label{eq:empcov_cont} \, .
\end{split}
\end{equation}
From the range of the sum, it can be seen that if $n,m<N$, then \opl{\Sigma}{nm} is computable using as most the first $2N$ terms of the series expansion.

The triple-integral is expressible in terms of the coefficients of the orthonormal exponentials as
\begin{equation}
  \opl{I}{ijk} = \sqrt{8}\sum\limits_{a=0}^i\sum\limits_{b=0}^j\sum\limits_{c=0}^k \frac{\opl{d}{ia}\opl{d}{jb}\opl{d}{kc} }{a+b+c}\, .
\end{equation}
There appears to be no simpler closed-form solution for \op{I}, but as it is a mathematical constant it need be calculated only once.

\subsubsection{Covariance matrix from the recursion relations}
The calculation of the covariance matrix via Eq.~\ref{eq:empcov_cont} is straightforward and useful, but requires $\mathcal O\left(N^2\Nmom\right)$ operations to calculate the $N\times N$ covariance matrix from \Nmom moments.  An $\mathcal O\left(N^2\right)$ algorithm also exists, which is especially useful for large covariance matrices computed from many moments.

Consider the first term in Eq.~\ref{eq:covdef}:
\begin{equation}
  \opl{f}{nm} = \intdz f\ofz E_n\ofz E_m\ofz \, .
\end{equation}
Note that this is written \op{f} to reflect the fact that it is a Hilbert-space operator corresponding to multiplication by $f\ofz$. This is distinct from the vector form \vec{f}.

Apply the recursion relations from Eq.~\ref{eq:recurrence} to $E_m\ofz$.  The result is
\begin{equation}
\begin{split}
  \opl{f}{nm+1} &= \intdz \frac{f\ofz E_n\ofz}{\phi_{2m+1}}\left(
                                4e^{-z}\Eb{m}\ofz 
                              - \frac{2}{\phi^2_{2m}}\Eb{m}\ofz 
                              - \phi_{2m-1} \Eb{m-1}\ofz
                   \right) \\
                &= \frac{1}{\phi_{2m+1}}\left(
                                4\intdz f\ofz e^{-z} E_n\ofz\Eb{m}\ofz 
                              - \frac{2}{\phi^2_{2m}} \opl{f}{nm}
                              - \phi_{2m-1} \opl{f}{nm-1}
                   \right) \\
                &= \frac{1}{\phi_{2m+1}}\left(
                                4 \oplu{e}{n}{~i} \opl{f}{in}
                              - \frac{2}{\phi^2_{2m}} \opl{f}{nm}
                              - \phi_{2m-1} \opl{f}{nm-1}
                   \right) \, ,
\end{split}
\end{equation}
where
\begin{equation}
  \opl{e}{ni} = \intdz e^{-z} E_n\ofz E_m\ofz
\end{equation}
is the operator representation of $e^{-z}$.  This matrix is tridiagonal and constant.  The elements can be obtained from Eq.~\ref{eq:recurrence}, and take the values
\begin{equation}
\begin{split}
  \opl{e}{nn}     &= \frac{1}{2} \phi_{2n}^{-2} \\
  \opl{e}{nn\pm1} &= \frac{1}{4} \phi_{2n\pm1} \, .
\end{split}
\end{equation}
The covariance matrix can therefore be obtained as
\begin{equation}
  \opl{\Sigma}{nm} = \opl{f}{nm} - \vecl{f}{n}\vecl{f}{m} \,
\end{equation}
where \op{f} is calculated either as $\opl{f}{nm} = \vecu{f}{i}\opl{I}{inm}$ or by using the recursion relations above.

\subsection{Optimal signal estimators}\label{sec:sigest}
Here we construct optimal estimators for a distribution's resonant contributions.  We use a construction that generalizes the concept of the Best Linear Unbiased Estimator (BLUE) to the problem of simultaneously estimating several parameters.

Suppose that some function $F\ofz$ is to be modeled as a linear combination of $N$ functions $f_1\dots f_N$:
\begin{equation}
  F\ofz = c^{\I m} f_{\I m}\ofz \, .
\end{equation}
We wish to construct a set of $N$ functions $\left\{\vecl{\omega}{\I n}\right\}$ such that
\begin{align}
  \inner{\omega_{\I n}}{f_{\I m}} = \delta_{nm} \, ,
\end{align}
that is, if \vecl{\omega}{\I n} is applied to $F\ofz$, its expected value is $c_\I{n}$.  These functions have covariance $\sigma_\I{nm}$, given by
\begin{align}
  \sigma_\I{nm} &= \intdz F\ofz \omega_{\I n}\ofz \omega_{\I m}\ofz - \left[\intdz F\ofz\omega_{\I n}\ofz\right] \left[\intdz F\ofz\omega_{\I m}\ofz\right] \\
                &= \veclu{\omega}{\I n}{~~i} \opl{\Sigma}{ij} \veclu{\omega}{\I m}{~~j}
\end{align}
where \op{\Sigma} is the covariance matrix associated with $F\ofz$.

We call the set $\left\{\vecl{\omega}{\I n}\right\}$ \textit{optimal} if the entropy of $\sigma$ is a minimum with respect to the set of all possible linear, unbiased estimators.  That is, we minimize $H=\ln\det\left(2\pi e\sigma\right)$.  This can be pictured as minimizing the volume of the $N$-dimensional ellipsoid described by $\sigma$.

This can be accomplished by introducing $N^2$ Lagrange multipliers $\eta_\I{ij}$ to produce a new objective function,
\begin{equation}
  \mathcal L = \ln\det\left(2\pi e\sigma\right) - \eta^\I{nm} \left(\vecl{\omega}{\I ni}\veclu{f}{\I m}{~~i} - \delta_{nm} \right) \, ,
\end{equation}
where the constraints enforce linearity and unbiasedness.  Without loss of generality, take $\sigma$ to be diagonal.  Then the objective function and its derivatives can be written
\begin{align}
  \mathcal L &= \sum\limits_{n=0}^N\ln\left(2\pi e\sigma_\I{nn}\right) - \eta^\I{nm} \left(\vecl{\omega}{\I ni}\veclu{f}{\I m}{~~i} - \delta_{nm} \right) \\
  \frac{d\mathcal L}{d\vecl{\omega}{\I ni}}   &= \frac{2}{\sigma_\I{nn}} \opl{\Sigma}{ij}\veclu{\omega}{\I n}{~~j} - \eta_\I{n~}^\I{~m} \vecl{f}{\I mi} \label{eq:LagrangeEpsilon} \\
  \frac{d\mathcal L}{d\eta_\I{nm}} &= \vecl{\omega}{\I ni}\veclu{f}{\I m}{~~i} - \delta_{nm} \, .
\end{align}
Setting Eq.~\ref{eq:LagrangeEpsilon} to zero and absorbing factors of $\frac{1}{2}\sigma_\I{nn}$ into $\eta^\I{nm}$,
\begin{equation}
  \vecl{\omega}{\I ni} = \eta_\I{n~}^\I{~m} \oplu{\Sigma}{ij}{-1} \veclu{f}{\I m}{~~j} \, .
\end{equation}
Take the dot product with \veclu{f}{\I l}{~~n}, and then multiply both sides by $\eta^{-1}_\I{ik}$, yielding
\begin{equation}
  \eta^{-1}_\I{nm} = \veclu{f}{\I n}{~~i} \oplu{\Sigma}{ij}{-1} \veclu{f}{\I m}{~~j} \, .
\end{equation}
Note that here, raising an object to the power minus-one indicates the matrix inverse.

We call $\vecl{\epsilon}{\I n} = \oplu{\Sigma}{ij}{-1} \veclu{f}{\I n}{~~j}$ the \textit{minimum variance estimator} for $f_n$ and $\sigma_\I{nm}$ the orthogonalization matrix for the functions $f_1,\dots,f_n$.

\end{document}